\documentclass[twocolumn,twocolappendix]{aastex631}
\usepackage{CJK}

\newcommand{\mz}[1]{{#1}}
\newcommand{\mztwo}[1]{{#1}}
\newcommand{\mzac}[1]{{#1}}
\shorttitle{Distances to local clouds}
\shortauthors{Zhang}
\graphicspath{{./}{figures/}{figures/figureset/}}

\begin{document}
\begin{CJK*}{UTF8}{gbsn}

\title{Distances to nearby molecular clouds traced by young stars}

\author[0000-0002-6388-649X]{Miaomiao Zhang (张淼淼)}
\affiliation{Purple Mountain Observatory, and Key Laboratory for Radio Astronomy, Chinese Academy of Sciences, Nanjing 210023, PR China}









\begin{abstract}
I present a catalog of distances to \mz{63} molecular clouds located within $\sim$2.5 kpc of the Sun. The cloud distances are derived based on utilizing the {\it Gaia} \mz{DR3 parallaxes} of the young stellar objects (YSOs). By identifying AllWISE YSO candidates (YSOCs) with infrared excesses and combining them with published YSOC catalogs, I compile an all-sky YSOC sample that is devoid of a significant proportion of contaminants. Using {\it Gaia} \mz{DR3 astrometric} measurements, I associate over \mz{3000} YSOCs with \mz{63} local clouds and obtain the average distance to each cloud by fitting the YSOC \mz{parallax} distribution within the cloud. I find good agreements with typical scatter of $\lesssim$10\% between my new cloud distances and previous distance estimates. Unlike cloud distances obtained using stellar extinction, my catalog provides distances to the relatively dense areas of local clouds, which makes them more appropriate references for investigating the physical properties of nearby dense regions.

\end{abstract}




\section{Introduction} \label{sec:intro}
In order to understand the process of star formation in galaxies, it is essential to have a comprehensive understanding of how molecular gas is transformed into stars in our local environment. To gain insight into this process within our solar neighborhood, it is necessary to investigate the structures and distributions of local clouds.
However, the gas structures within $\sim$2 kpc of the Sun is still under debate \citep{poppel2000,bouy2015,zari2018,zuckernature2022}, particularly in light of recent discoveries of several new structures \citep{radcliffe2020,bialy2021,mcbride2021}.

Accurately mapping the three-dimensional (3D) gas structures in our solar neighborhood is contingent upon precise distance measurements to nearby molecular clouds. 
Several methods have been developed to determine the distances of these clouds. For instance, the statistics of star counts on obscured and non-obscured fields can provide insights into cloud distances 
\citep{wolf1923,bok1931,lombardi2010,kainulainen2011,foster2012}. 
However, this approach often suffers from large uncertainties due to the uneven distribution of stellar density. Photometric distances of cloud tracers, such as OB associations, can also be utilized to estimate cloud distance 
\citep{borgman1964,garrison1967,brown1994,mayne2008}, 
but the inherent degeneracy between distance and extinction of stars often results in significant inaccuracies.

Another effective method for estimating cloud distances is by using the trigonometric parallaxes of masers or YSOs located within the clouds. For instance, \citet{de1999} examined nearby OB associations based on HIPPARCOS \citep{hipparcos1997} data and obtained distance estimates for tens of these associations. However, the precision of HIPPARCOS parallax measurements only allows for visits to clouds located within distances of up to 600 pc. The Gould's Belt Distances Survey \citep[BOBELINS,][]{gobelins2011,loinard2013} utilized radio very long baseline interferometry (VLBI) observations to obtain parallaxes for a large sample of young stars in various nearby molecular clouds. These high-precision parallax measurements enable accurate distance determinations for individual sources. By averaging the distances of young stars, GOBELINS derived the distances to clouds such as Ophiuchus, Orion, Serpens, Taurus, and Perseus \citep{gobelins-oph,gobelins-orion,gobelins-serpens,gobelins-taurus,gobelins-perseus}. Nevertheless, VLBI observations are time-consuming,  and therefore the number of sources with VLBI parallaxes is limited, which restricts the application of GOBELINS's method to a small number of clouds. 

The recent release of astrometric data from {\it Gaia} has brought about significant changes. {\it Gaia} DR2 \citep{gaiadr2} \mz{and DR3 \citep{gaiadr3-summary} have provided} accurate parallax measurements for more than one billion sources,  \mz{prompting numerous authors to re-calculate the distances to nearby molecular clouds using {\it Gaia} data. One approach involves identifying a stellar extinction ``jump" caused by molecular clouds between the unreddened foreground stars and the reddened background stars by examining the variation of optical extinction with respect to distance along the line of sight \citep{knude1998}. Hundreds of molecular clouds have had distances derived using this method  \citep{schlafly2014,yan2019,zucker2019,zucker2020}. }
\mz{Another approach involves constructing a 3D extinction map by modeling the dust extinction profiles along different lines of sight \citep{rezaei2018,chen2019,green2019,lallement2019,leike2020}. Cloud distances can then be obtained by searching for dust structures in the 3D dust cubes \citep{chen2020,guo2022}. \mz{These techniques rely on estimates of stellar extinction, but the dynamical range of stellar extinction based on optical surveys such as {\it Gaia} is limited, only extending up to a few to $\sim$10 magnitudes ($A_V$). As a result, extinction-based techniques are not applicable for estimating distances of relatively dense regions.}}

The distances of nearby clouds can also be obtained from {\it Gaia} parallaxes of YSOs. Several studies have identified a significant number of YSOCs using {\it Gaia} photometry and/or kinematics \citep{zari2018,banyan11,banyan12,banyan13,kounkel2019,kounkel2020,mcbride2021,prisinzano2022}. While \citet{zari2018} and \citet{mcbride2021} have analyzed the spatial distribution of YSOCs, their focus has mainly been on structures within $\sim$500 pc of the Sun. \citet{prisinzano2022} used machine learning unsupervised clustering algorithm DBSCAN \citep[Density-Based Spatial Clustering of Applications with Noise,][]{dbscan1996} to identify 354 star forming regions and stellar clusters with ages of $\lesssim$~10 Myr within $\sim$1.5 kpc of the Sun based on their {\it Gaia} YSO sample. However, the star forming regions identified by \citet{prisinzano2022} are stellar structures, and the connections between them and gas structures are still uncertain. On the other hand, \citet{dzib2018} has compiled a YSO catalog from the literature for 12 nearby clouds within 500 pc of the Sun. Using {\it Gaia} parallax measurements of these YSOs, \citet{dzib2018} investigated the distances and 3D motions of these 12 local clouds. 

In this paper, I expand on the distance estimation of local clouds within $\sim$2.5 kpc of the Sun using {\it Gaia} DR3 parallax measurements. 
I compile a comprehensive all-sky sample of YSOCs and use it to 
precisely determine distances to 63 nearby molecular clouds. In Section~\ref{sec:catalogs}, I provide a detailed description of the data and catalogs utilized, as well as the quality cuts applied to \mz{filter out spurious sources.} 
In Section~\ref{sec:ysocat}, I explain how I constructed my all-sky YSOC sample, which is free of contamination from other types of sources. 
In Section~\ref{sec:results}, I present my method for determining cloud distances and provide the resulting distance catalog. 
In Section~\ref{sec:discussion}, I offer a comparison of my distance estimates with those from previous literature, highlighting the advantages and drawbacks of my approach. \mz{Finally, my conclusions and summary are presented in Section~\ref{sec:summary}}.

\section{Data and catalogs} \label{sec:catalogs}
I utilize the archival AllWISE catalog to identify YSOCs, and the kinematic information for these YSOCs is provided by the {\it Gaia} \mz{DR3}. In addition, I incorporate several previously published catalogs to augment my YSOC sample and provide distance estimates for the YSOCs. The nearby molecular clouds are delineated using the Planck dust maps. The succeeding sections provide a detailed account of these data and catalogs.

\subsection{AllWISE catalog} \label{sec:allwisecat}
The space telescope WISE was launched in December 2009 and scanned the whole sky in four infrared passbands, W1, W2, W3, and W4, centered at 3.4, 4.6, 12, and 22\,\micron, respectively. The angular resolution is about 6\farcs1, 6\farcs4, 6\farcs5, and 12\farcs0 in W1-W4 bands and the 5$\sigma$ sensitivity is better than 0.08, 0.11, 1, and 6 mJy in unconfused regions. I use the source catalog taken from  AllWISE data release\footnote{\url{https://wise2.ipac.caltech.edu/docs/release/allwise/}} \citep{allwisecat}, which was produced by combining WISE data from cryogenic and post-cryogenic survey phases. The detailed description of WISE data acquisition and reduction can be found in \citet{wise}, \citet{wisecali}, and Explanatory Supplement to the AllWISE Data Release Products\footnote{\url{https://wise2.ipac.caltech.edu/docs/release/allwise/expsup/}}. The AllWISE catalog also provides $J$, $H$, $K_s$ photometry by positionally crossmatching with the 2MASS point source catalog \citep{2mass}.

\subsection{{\it Gaia} \mz{DR3}} \label{sec:gaiadr2cat}
The \mz{third} release of the {\it Gaia} data\footnote{\mzac{\url{https://gea.esac.esa.int/archive/}}} (\mz{DR3}) is based on the data collected during the first \mz{34} months of {\it Gaia} mission \citep{gaiamission}. It provides high-precision parallax and proper motion, together with homogeneous multi-band photometry for \mz{about 1.8} billion sources. \mz{Actually, {\it Gaia} DR3 released a vast array of data products, including the spectra, photometric time series, and the astrophysical \mztwo{parameters. }
In this paper, I only use the astrometry, photometry, and astrophysical parameter catalogs in {\it Gaia} DR3.} \mz{I also note that the astrometry and broad band photometry in {\it Gaia} DR3 are the same\footnote{\mztwo{According to \citet{gaiaedr3}, a correction must be made to the G-band magnitudes of some of the sources in {\it Gaia} EDR3. This correction is not included in the official EDR3, but it is already incorporated in the official DR3. As a result, technically, the G-band photometry of certain sources in DR3 differs from that in EDR3.}} as that in {\it Gaia} EDR3 \citep{gaiaedr3} which was the first instalment of the full Gaia DR3} 

{\it Gaia} \mz{DR3} has a limiting magnitude of $G\approx$~21 mag and a typical uncertainty of 0.3 millimagnitude (mmag) ($G<$~13 mag) to \mz{6} mmag ($G=$~20 mag). The typical astrometric uncertainty depends on source brightness: \mz{0.02$-$0.04 milliarcseconds (mas) ($G<$~15 mag) to 0.5 mas ($G=$~20 mag) for parallax and 0.02-0.04 mas\,yr$^{-1}$ ($G<$~15 mag) to 0.6 mas\,yr$^{-1}$ ($G=$~20 mag) for proper motion.} 
\mz{The {\it Gaia} DR3 astrophysical products were produced with 13 different modules using the astrophysical parameters inference system
\citep[Apsis,][]{apsis2022,apsis-para2022}. In this paper, I use the astrophysical parameter catalog produced by one of 13 modules in Apsis, i.e., General Stellar Parameterizer from Photometry \citep[GSP-Phot,][]{gsp-phot2022}, based on the {\it Gaia} astrometry, photometry, and low-resolution BP/RP spectra. The GSP-Phot is a  Bayesian forward-modelling approach, which provides a homogeneous catalogue of stellar parameters, distances, and extinctions for about 471 million sources with $G<$~19 mag. \citet{gsp-phot2022} estimated that the typical uncertainty of extinction ($A_G$) is about 0.06 mag for bright sources.} The detailed content of {\it Gaia} \mz{DR3} can be found in \citet{gaiadr3-summary,gaiadr3-validation}.

\mz{To remove the spurious astrometric solutions from the {\it Gaia} DR3 catalog, I use the classifier introduced by \citet{rybizki2022}. \citet{rybizki2022} constructed the ``good" and ``bad" astrometric solutions as the training samples with the {\it Gaia} EDR3 data and then devised  a single ``astrometric fidelity" parameter to identify spurious sources based on the machine learning technique. Comparing with the quality cuts using the {\it Gaia} catalog columns such as $\mathtt{ruwe}$, their astrometric fidelities can yield purer and more complete samples of sources with reliable astrometric solutions. \citet{rybizki2022} also provided the diagnostics of the level of photometric contamination from neighbors ($\mathtt{norm\_dg}$). I decide to use the following cuts to filter out the spurious sources with unreliable astrometric solutions and/or colors in {\it Gaia} DR3 as suggested by \citet{rybizki2022}:}
\mz{\begin{eqnarray}
\mathtt{fidelity\_v2}&>&0.5,\\
\mathtt{norm\_dg}=\mathtt{nan} ~&or&~ \mathtt{norm\_dg}<-3
\end{eqnarray}
}
\mztwo{To remove the {\it Gaia} sources with possible problematic photometries, \mz{I also apply a quality cut as suggested by \citet{prisinzano2022}}:
\begin{eqnarray}
    \sigma([G-R_p])=\sqrt{\sigma(G)^2+\sigma(R_p)^2}<0.14
\end{eqnarray}
\mz{where $\sigma(G)$ and $\sigma(R_p)$ are defined as}:
\begin{eqnarray*}
\sigma(G)&=&\sqrt{\left( \frac{1.0857}{\mathtt{phot\_g\_mean\_flux\_over\_error}} \right)^2+\sigma(G_0)^2}\\
\sigma(R_p)&=&\sqrt{\left( \frac{1.0857}{\mathtt{phot\_rp\_mean\_flux\_over\_error}} \right)^2+\sigma(R_{p0})^2}\\
\end{eqnarray*}
and $\sigma(G_0)$ and $\sigma(R_{p0})$ are the {\it Gaia} DR3 zero-point uncertainties\footnote{\url{https://www.cosmos.esa.int/web/gaia/dr3-passbands}}.}

\mz{Finall, I correct the {\it Gaia} DR3 parallax using the zero point bias suggested by \citet{gaiaparallaxbias2021}, which is a function of source position, brightness, and color\footnote{\url{https://gitlab.com/icc-ub/public/gaiadr3_zeropoint}}}.

\subsection{Planck dust map} \label{sec:plankdustmap}
The Planck mission \citep{planckmission2011} mapped the whole sky in nine passbands in the range of frequencies between 25 and 1000\,GHz. 
Based on the 2013 release of data \citep{planck2013release}, \citet{planckdustmap2014} fit the emission from Planck data at 353, 545, 857\,GHz, and IRAS 100\,\micron~survey \citep{irasmission} data using an all-sky dust model that describes the dust spectral energy distribution (SED) with a modified blackbody. The released dust opacity ($\tau$) map has a angular resolution of 5\arcmin~and was also calibrated to the reddening $E(B-V)$ units with extragalactic quasars. I fetch Planck $E(B-V)$ map using the Python interface of DUSTMAPS \citep{dustmapspython} and transform it to visual extinction units with $A_V=$~3.1$E(B-V)$ assuming $R_V=$~3.1.

\subsection{Complementary published catalogs} \label{complementarycats}
\subsubsection{\citet{ysowise2016,ysowise2019}}\label{sec:comp-ysowise}
\citet{ysowise2016} identified 133\,980 Class I/II candidates and 608\,606 Class III/evolved YSOCs based on the AllWISE catalog. More specifically, they collected known sources from SIMBAD\footnote{\url{http://simbad.u-strasbg.fr/simbad/}} as the training sample and then classified AllWISE sources into different types with the support vector machine (SVM) technique. After excluding the contaminations such as extragalactic sources, main-sequence \mz{(MS)} stars, and evolved stars, \citet{ysowise2016} finally selected 742\,586 YSOCs. 

\citet{ysowise2019} combined {\it Gaia} DR2 and AllWISE catalog and identified 1\,768\,628 potential YSOCs with the supervised machine learning technique. Their training sample was constructed based on SIMBAD and $\sim$80 catalogs from the literature. After comparing the results from tens of different machine learning techniques, \citet{ysowise2019} finally selected the Random Forests method to classify {\it Gaia} DR2$\times$AllWISE sources into different object classes, e.g., YSOCs and contamination such as \mz{MS} stars and extragalactic objects.

\subsubsection{\mz{\citet{gaiadist2021}}} \label{sec:comp-gaiadist}
The transformation from {\it Gaia} \mz{DR3} parallax to distance needs to account for the non-gaussian profile of probability distribution of the inverse of parallax. \citet{gaiadist2021} calculated the Bayesian distances for about \mz{1.3-1.5} billion {\it Gaia} sources by assuming a prior \mz{constructed from a
3D Galactic model based on {\it Gaia} EDR3 data}. Their catalog can provide meaningful distance estimates even for the faint {\it Gaia} stars with large parallax uncertainties. The distance catalog presented by \mz{\citet{gaiadist2021}} has been included in the official {\it Gaia} data archive\footnote{\url{https://gea.esac.esa.int/archive/}}.

\mz{
\citet{gaiadist2021} provided two types of distance: the geometric distance ($\mathtt{r\_med\_geo}$) was obtained using parallax with a direction-dependent prior on distance; and the photogeometric distance ($\mathtt{r\_med\_photogeo}$) was obtained by considering additional stellar photometric information. In this paper, I only use the geometric distance in order to get the distances for more {\it Gaia} sources. I also emphasize that the \citet{gaiadist2021}'s geometric distances are only used to estimate the extinctions of some YSOCs (see Appendix~\ref{ap:extalphac}) and to exclude the possible contamination (see Sect.~\ref{sec:contamex}). I do not use them to calculate the final distances of YSOCs in the nearby molecular clouds.
}
    .

\section{All-sky YSOC catalog} \label{sec:ysocat}
In this section, I describe the methodology used to identify YSOCs in Section~\ref{sec:ysoiden}, the integration of previously published YSOC catalogs in Section~\ref{sec:ysocom}, the de-reddening and classification of YSOCs in Section~\ref{sec:ysoc-deredden}, and the removal of potential contamination sources in Section~\ref{sec:contamex}. The resulting clean YSOC catalog, along with estimates of contamination and completeness fractions, are presented in Section~\ref{sec:cleanyso}.

\subsection{YSOC identification} \label{sec:ysoiden}
The excessive infrared emission from YSOs can be used to distinguish them from field stars. In this paper, I use the multicolor criteria scheme suggested by \citet{koenig2014} to identify YSOCs with infrared excess based on the AllWISE catalog. The details about this multiphase source classification scheme can be found in \citet{koenig2014}. Here I just give a short description of this process.

\citet{koenig2014} found that the spurious detection fraction of AllWISE sources can be up to 30\% in the W1 band and $>$90\% in the W3 and W4 bands. To eliminate the spurious detections, they developed some criteria based on the signal-to-noise and reduced chi-squared parameters given in the AllWISE catalog, which can suppress the contamination rate down to $<$7\% in any WISE band.

\begin{figure*}
\includegraphics[width=1.0\linewidth]{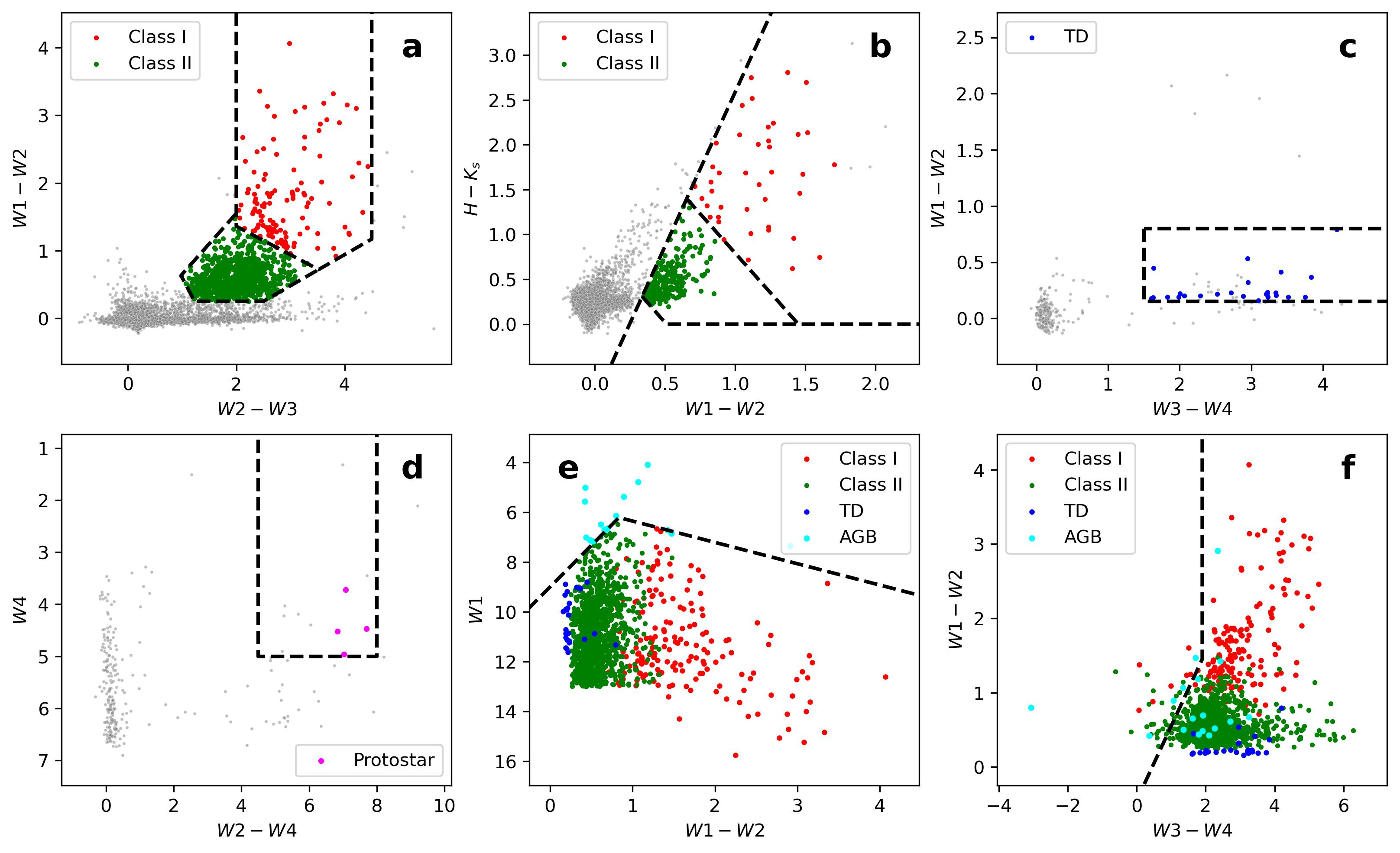}
\caption{Multicolor criteria scheme used to identify YSOCs from AllWISE catalog for a example of a 11\degr$\times$4\degr~region towards the Orion A. The gray dots show the distribution of field stars in the color space. The protostars and Class I candidates are marked with magenta and red circles, respectively. The green dots label Class II candidates while the blue circles mark transition disks. The AGB candidates isolated from YSOCs are labelled with cyan circles. \label{fig:ysoiden}}
\end{figure*}

After filtering out the spurious detections, I firstly remove the contamination of star-forming galaxies and Active Galactic Nuclei (AGNs) based on their photometry in W1, W2, and W3 bands. Then the Class I and Class II candidates are selected using the color criteria shown in Fig.~\ref{fig:ysoiden}a. Secondly, the YSOC with $H$ and $K_s$ detections are identified from remaining sources with the color criteria shown in Fig.~\ref{fig:ysoiden}b. Thirdly, by introducing W4 photometry, transition disks (TDs) are identified from the remaining sources (see Fig.~\ref{fig:ysoiden}c) while the protostars are retrieved from AGN candidates (see Fig.~\ref{fig:ysoiden}d). Finally, all identified YSOCs mentioned above are reexamined to isolate the possible asymptotic giant branch (AGB) stars with the color criteria shown in Fig.~\ref{fig:ysoiden}e and f.

I ultimately obtain 107\,401 YSOCs in the whole sky, including 20\,317 Class I candidates, 59\,543 Class II candidates, 1650 TDs, 141 protostars, and 25\,750 AGB candidates.
\mz{I note that the number of YSOCs (107\,401) obtained with above criteria is smaller than that (133\,980) presented by \citet[][]{ysowise2016} which is also based on AllWISE data, but obtained with a machine-learning method (see Sect.~\ref{sec:comp-ysowise}). \citet{ysowise2016} compared their results with that derived by \citet{koenig2014}'s color criteria: their SVM method is more successful in excluding extragalactic contamination and able to recover higher fraction of known YSOs; \citet{koenig2014}'s method is more sensitive to fainter sources and thus efficient in identifying the Galactic contamination, but retrieve a lower fraction of known YSOs. Therefore, combining these catalogs obtained with different methods could lead to a more complete YSOC sample.}

\subsection{Assembling a combined YSOC catalog}\label{sec:ysocom}

I cross-match the YSOCs identified in Sect.~\ref{sec:ysoiden} and the published YSOC catalogs by \citet{ysowise2016,ysowise2019} (see Sect.~\ref{sec:comp-ysowise}) based on the AllWISE source ID and finally obtain a combined YSOC catalog that includes 2\,551\,895 YSOCs.

I have also conducted a cross-match between the combined YSOC catalog and {\it Gaia} \mz{DR3} in order to obtain parallax and proper motion information. The cross-matching procedure used is described in the {\it Gaia} DR3 documentation \mz{\citep{gaiadr3doc}}, which includes the matching of {\it Gaia} DR3 with various external survey catalogs, such as the AllWISE catalog, using an algorithm that takes into account source position errors, proper motions, and environment \mz{\citep{gaiacrossmatch2017,gaiaxwise2019}}. The matched catalogs that provide the {\it Gaia} source IDs and the corresponding external catalog source IDs are included as part of the official {\it Gaia} \mz{DR3}. Initially, I cross-matched our combined YSOC catalog with the BestNeighbour table by \mz{\citet{gaiadr3doc}} based on AllWISE source ID, and then retrieved {\it Gaia} \mz{DR3} entries using the obtained {\it Gaia} \mz{DR3} source ID. It should be noted that some YSOCs have multiple {\it Gaia} \mz{DR3} counterparts, and for these sources, I only retained the closest match. Additionally, I cross-matched the combined YSOC catalog with the distance catalog by \citet{gaiadist2021} (as described in Sect.~\ref{sec:comp-gaiadist}) based on the {\it Gaia} DR3 source ID.


The final combined catalog includes over two million YSOCs, of which \mz{$\sim$68\% has {\it Gaia} DR3 counterparts and $\sim$36\% has GSP-Phot extinction estimates. I also note that all {\it Gaia} DR3 counterparts have \citet{gaiadist2021}'s geometric distance estimates due to the removal of spurious sources (see Sect.~\ref{sec:gaiadr2cat}).}

\subsection{YSOC de-reddening and classification}\label{sec:ysoc-deredden}

To analyze the intrinsic properties of YSOs, it is necessary to correct for extinction in their fluxes. However, estimating the extinction towards individual YSOs is a non-trivial task. Approximately 36\% of my YSOCs already have GSP-Phot extinction estimates ($A_G$), which were obtained by modeling the {\it Gaia} BP/RP spectrum, parallax, and apparent $G$ magnitude using stellar evolutionary models \citep[PARSEC,][]{stellarmodel_parsec2012} and several synthetic spectra libraries \citep{gsp-phot2022}. It should be noted that the stellar evolutionary models do not include tracks for YSOs with surrounding disks \citep{stellarmodel_parsec2012}, which means that the $A_G$ values for protostars with significant infrared excess may not be reliable. As a result, I need to classify my YSOCs into different categories based on their infrared excess and recalculate their foreground extinction separately.

The YSO classification is usually based on the spectral index that is defined as
\begin{equation}
    \alpha = \frac{d\mathrm{log}(\lambda S_{\lambda})}{d\mathrm{log}(\lambda)},
\end{equation}
where $S_{\lambda}$ is the flux density at wavelength $\lambda$. By fitting the de-reddened SEDs from 2 to 22\,\micron, the YSOs can be classified as Class I, Class II, and \mz{Class III} sources based on the scheme suggested by \citet{lada1987}. Therefore, the reliable YSO classification should be performed after the YSO de-reddening, which results in the coupling of YSO de-reddening and classification.

I finally decide to use a iterative process to do the de-reddening and classification for my YSOCs, which can obtain the foreground extinctions \mz{($A_{G,\mathrm{final}}$)} and de-reddened spectral indices ($\alpha_c$) of the YSOCs at once. The idea is to estimate the extinctions of YSOCs based on their classification and then to re-classify the YSOCs using $\alpha_c$, iteratively, until $\alpha_c$ approaches constants. The detailed steps of the process are described in Appendix~\ref{ap:extalphac}. Figure.~\ref{fig:hist_alpha} shows the distributions of the observed spectral indices ($\alpha$) and $\alpha_c$ of YSOCs. Based on $\alpha_c$, \mztwo{of over two million YSOCs}, I finally obtain \mztwo{$\sim$3\% }
Class I candidates, \mztwo{$\sim$16\% }
Class II candidates, and \mztwo{$\sim$81\% }
Class \mz{III} candidates using the classification scheme from \citet{lada1987}.

\begin{figure*}
    \centering
    \includegraphics[width=1.0\linewidth]{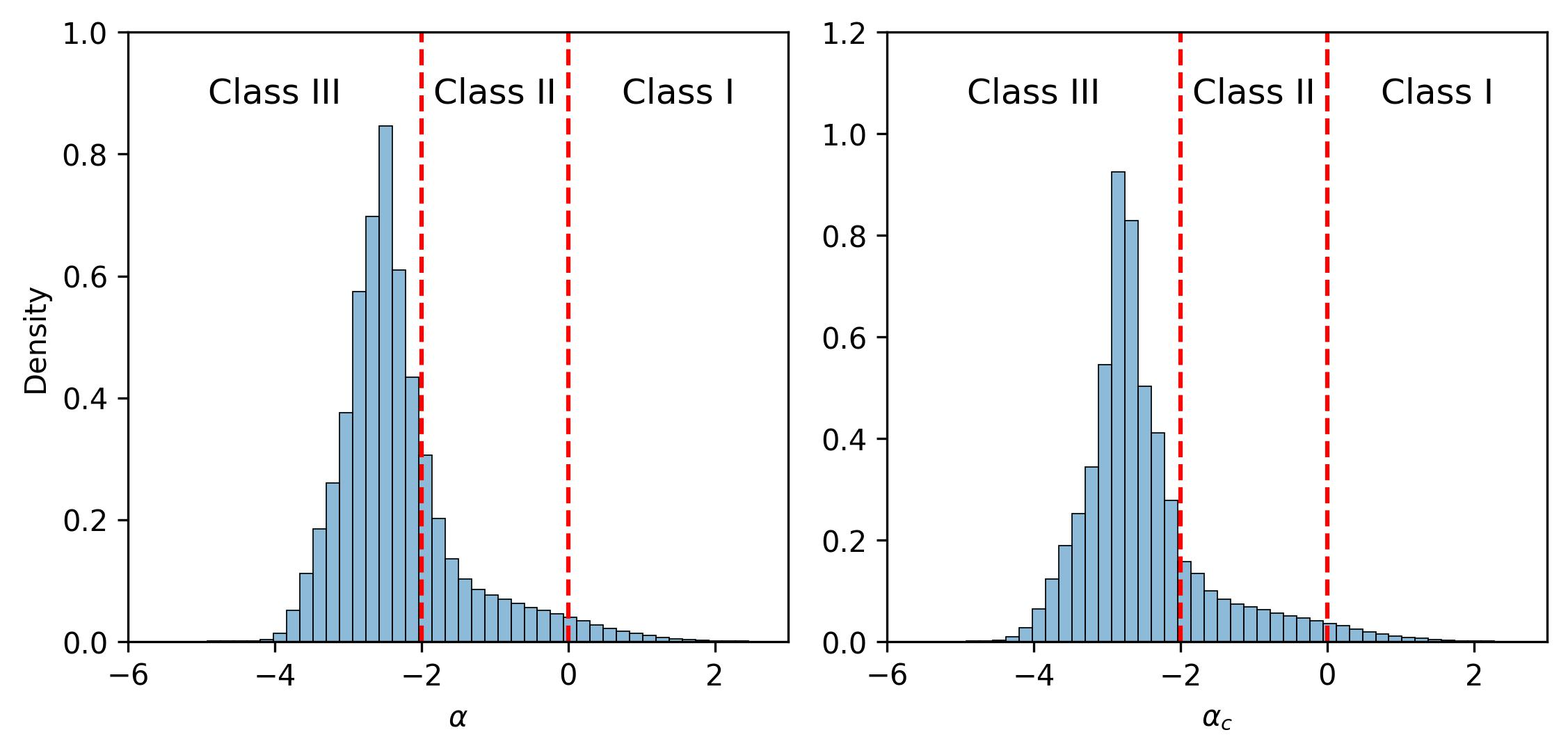}
    \caption{Histograms of the observed spectral indices ($\alpha$, left panel) and the de-reddened spectral indices ($\alpha_c$, right panel) of YSOCs. The red dashed lines mark the criteria of the YSO classification scheme suggested by \citet{lada1987}.}
    \label{fig:hist_alpha}
\end{figure*}

\subsection{Estimating and excluding possible contamination}\label{sec:contamex}

My combined YSOC sample is likely contaminated by different kinds of objects, including the MS stars, AGBs, red giants, and extragalactic sources. 

A large number of extragalactic sources have been mitigated during the YSO identification process described in Sect.~\ref{sec:comp-ysowise} and \ref{sec:ysoiden}. However, there could be galaxies remaining in the final YSOC catalog. I use the data from Spitzer Wide-Area Infrared Extragalactic Survey \citep[SWIRE,][]{swire2003} to estimate the residual contamination fraction of galaxies and AGNs in my YSOC catalog. SWIRE performed the IRAC and MIPS observations on six sky fields that covered about 65.6 deg$^2$ in total. I download the Spring'05 catalogs for the ELAIS N1, ELAIS N2, Lockman, and XMM\_LSS regions, and Fall'05 catalogs for the CDFS and ELAIS S1 regions \mz{\citep{swiredata}}. All detected SWIRE sources are cross-matched with my YSOCs using the tolarence of 3\arcsec. Finally I obtain 37 SWIRE extragalactic sources in my YSOC catalog, which results in a surface density of 37/65.6$\sim$0.56 extragalactic sources per deg$^2$. After scaling to the whole sky, there could be 23\,261 galaxies in my YSOC catalog if assuming that the extragalacitc sources distributed uniformly in the whole sky. Considering the Galactic extinction, this number should be the upper limit. Therefore, the contamination fraction of the extragalactic sources is negligible ($<$1\%) in my YSOC catalog.

In Sect.~\ref{sec:ysoiden}, I have already isolated a number of possible AGBs with the multicolor criteria. However, the dusty AGBs can also produce infrared excess, which means that it is difficult to distinguish AGBs from YSOs only based on the color criteria. \citet{ysowise2016,ysowise2019} classified a large number of AllWISE sources into the evolved stars that is a grouped type including many subtypes such as AGBs, post-AGBs, RGBs, and evolved supergiants. However, their training sample was collected from the SIMBAD and literature, including many candidates rather than the only bona-fide labeled objects. The noise in the training data would propagate to the noise in the predictions. Actually, \citet{mcbride2021} checked the parallaxes of \citet{ysowise2019}'s {\it Gaia}-AllWISE YSOC sample in some star forming regions such as Ophiuchus. They found that although the {\it Gaia}-AllWISE YSOCs recovered some sources associated with the star-forming regions the vast majority of YSOCs were just background sources. Therefore, the contamination fraction of MS stars, giants, and AGBs could be very high in my combined YSOC catalog.

To exclude the possible MS stars, giants, and AGBs, I introduce the PARSEC \citep{stellarmodel_parsec2012} and COLIBRI \citep{stellarmodel_colibri2013} stellar evolutionary tracks, and YSO models from \citet{ysomodel2006,ysomodel2007}. The PARSEC tracks were computed for different chemical compositions and evolutionary phases from pre-main sequence (PMS) till the onset thermally pulsing AGB (TP-AGB) while the COLIBRI tracks extended the evolution to the end of the TP-AGB phase. I use the web interface, CMD\footnote{\url{http://stev.oapd.inaf.it/cmd}} (version 3.4), to extract the isochrones of MS stars, giants, and AGBs with the age range of 0.01\,Myr$-$13.5\,Gyr and solar metallicity. \citet{ysomodel2006} presented a grid of radiation transfer models of YSOs, including about 200\,000 YSO models. The grid covered a wide range of stellar, disk, and envelope masses, and accretion rates. The SEDs of each YSO model were calculated assuming ten different inclination angles and then convolved to many commonly-used filters to produce broadband fluxes within 50 apertures of $\sim$100$-$100\,000\,AU. I select the YSO models with the inclination angles between 30\degr~and 60\degr, stellar masses between 0.08 and 10\,$M_{\odot}$, disk-to-stellar mass ratios between 0 and 1 and then extract the {\it Gaia} and WISE fluxes within the aperture of $\sim$45\,000\,AU. 

\begin{figure*}
    \centering
    \includegraphics[width=1.0\linewidth]{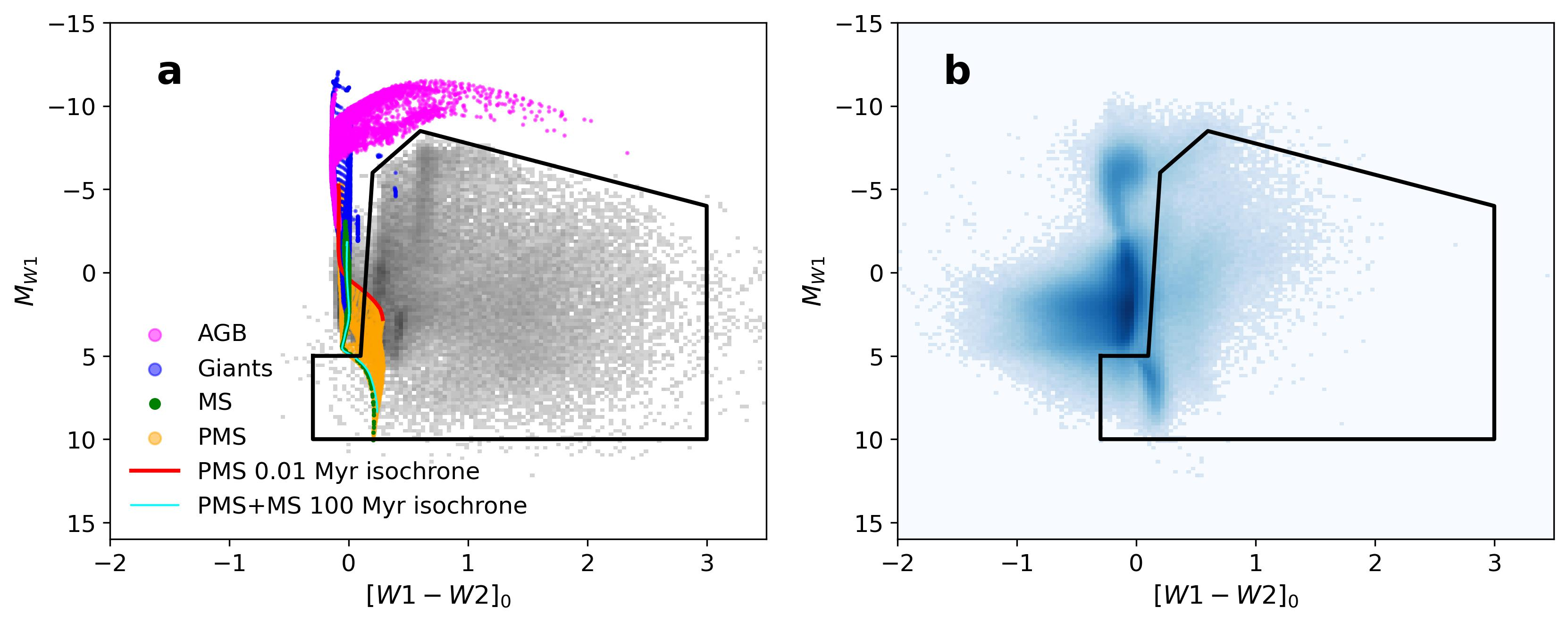}\\%
    \includegraphics[width=1.0\linewidth]{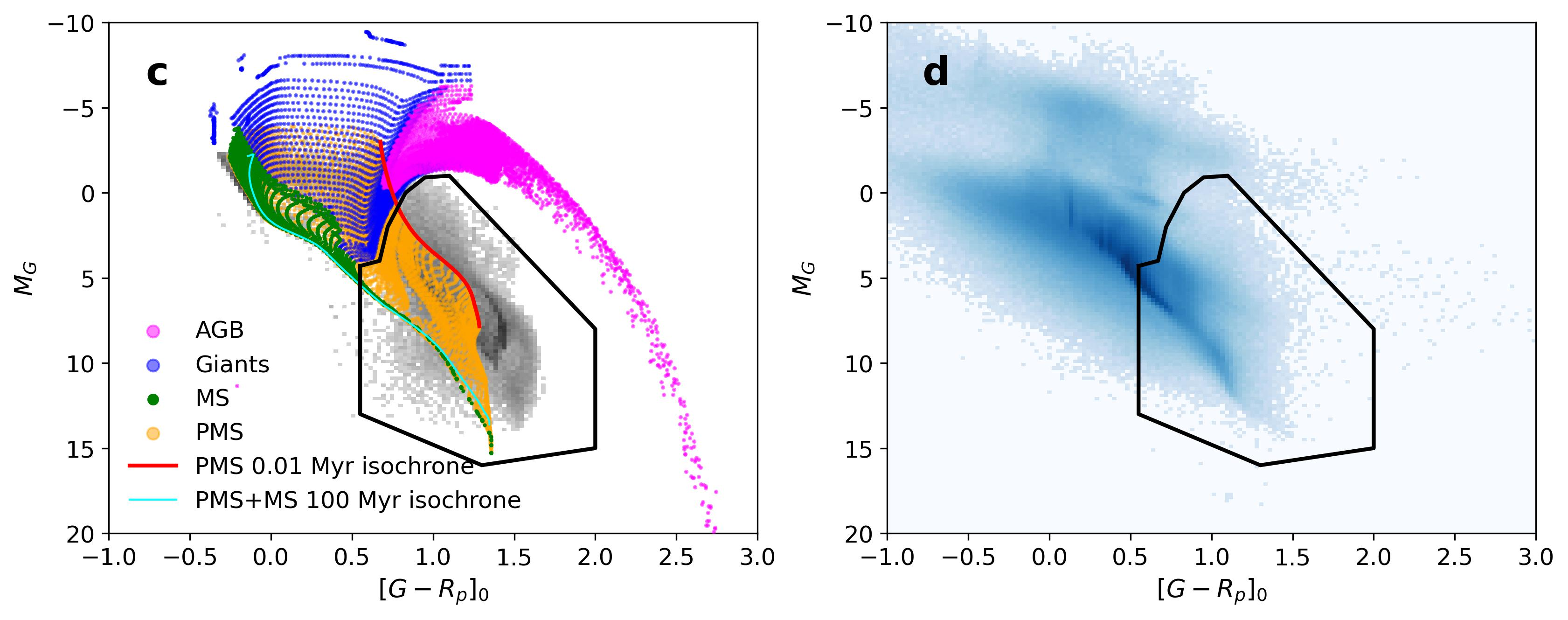}
    \caption{Criteria used to isolate YSOs from MS, giants, and AGBs. (a): CMD of the absolute magnitude in W1 band, $M_{W1}$, versus the intrinsic color of $[W1-W2]_0$ for the evolutionary tracks of PMS, MS, giants, AGBs \citep[][]{stellarmodel_parsec2012,stellarmodel_colibri2013}, and the YSO models \citep[][]{ysomodel2006}; (b): $M_{W1}$ versus $[W1-W2]_0$ CMD for the YSOCs with the photometric uncertainties of $\sigma(W1)<$\,0.2\,mag and $\sigma(W2)<$\,0.2\,mag; (c): CMD of the absolute magnitude in {\it Gaia} G band, $M_{G}$, versus the {\it Gaia} \mz{intrinsic color of $[G-R_p]_0$} for the evolutionary tracks of PMS, MS, giants, AGBs, and the YSO models; (d): \mz{$M_G$ versus $[G-R_p]_0$} CMD for the selected YSOCs (see text for details). The golden dots and green dots mark the PMS and MS tracks with the mass range of 0.08$-$10\,$M_{\odot}$, respectively. The blue dots label the giant tracks, including subgiant branch (SGB) and red giant branch (RGB). The early-AGBs, TP-AGBs, and post-AGBs are all marked with the magenta dots. The red curves show the PMS 0.01 Myr isochrones while the cyan curves are the joint isochrones for PMS and MS tracks with the age of 100 Myr. The YSO models from \citet{ysomodel2006} are shown with background gray density maps. The black solid polygons define the criteria and the YSOCs located outside the polygons are identified as contamination.}
    \label{fig:cmd_deredden_wise}
\end{figure*}

Figure~\ref{fig:cmd_deredden_wise} (left panels) shows the color-magnitude diagrams (CMDs) of the absolute magnitudes ($M_{W1}$ and $M_{G}$) versus the intrinsic colors ($[W1-W2]_0$ and \mz{$[G-R_p]_0$}) for the selected evolutionary tracks and YSO models mentioned above. \mz{The selection of color, $[G-R_p]$, is to avoid the overestimation of mean $B_p$ magnitude for the faint sources due to the application of the minimum flux threshold \citep{riello2021}.}
I define two polygons in the color spaces of $M_{W1}$ vs. $[W1-W2]_0$ and \mz{$M_G$ vs. $[G-R_p]_0$}, respectively. Table~\ref{tab:polygons} gives the vertex of polygons. 
I discovered that the majority of giants and AGBs are situated outside the polygons, indicating that the polygons' criteria can effectively eliminate the contamination from these sources. However, there are still some limitations to the polygons' use. Firstly, many bright PMS stars are excluded because they overlap with giants and AGBs in the color space. Secondly, despite the defined polygons removing most of the MS stars, some MS stars remain inside the polygons (as seen by the green dots in Fig~\ref{fig:cmd_deredden_wise}). As a result, using these polygons can produce a relatively clean but less complete YSO sample.

To apply the defined polygons to my YSOCs, I need to estimate their absolute magnitudes and intrinsic colors. First, I select \mztwo{about 2.2 million }
YSOCs that are detected in the WISE W1 and W2 band with the photometric uncertainties of $\sigma(W1)<$\,0.2 and $\sigma(W2)<$\,0.2\,mag and have \mz{\citet{gaiadist2021}'s geometric distance estimates} 
and positive \mz{extinction estimates ($A_{G,\mathrm{final}}$; see details in Appendix~\ref{ap:extalphac})}. Figure~\ref{fig:cmd_deredden_wise}b shows the $M_{W1}$ vs. $[W1-W2]_0$ CMD for the selected YSOCs, of which \mztwo{over 120 thousands }
YSOCs are located inside the polygon. Second, \mz{most of YSOCs \mztwo{inside $M_{W1}$ vs. $[W1-W2]_0$ polygon} have {\it Gaia} photometry ($G$ and $R_p$)}. 
Figure~\ref{fig:cmd_deredden_wise}d shows \mz{the $M_{G}$ vs. $[G-R_p]_0$ CMD} for the YSOCs with {\it Gaia} photometries, of which \mztwo{over 70 thousands }
YSOCs are located inside the polygon. Third, adding \mztwo{about 1000 }
YSOCs that are located inside the $M_{W1}$ vs. $[W1-W2]_0$ color space polygon but without {\it Gaia} photometries, I obtain \mztwo{about 78 thousands }
YSOCs as a relatively clean YSOC sample. Considering that the fraction of contamination can be higher than 50\% for Class III sources which were identified based on infrared excess \citep{oliveira2009,romero2012,dunham2015,manara2018}, I finally select \mz{24\,883 }
Class I/II YSOCs ($\alpha_c>-$2) to construct the final clean YSOC sample.

\begin{deluxetable}{cccc}
\tablecaption{\mz{Vertex of polygons in color space}\label{tab:polygons}}
\tablewidth{0pt}
\tablehead{
\colhead{$M_{W1}$} & \colhead{$[W1-W2]_0$} & \colhead{$M_G$} & \colhead{\mz{$[G-R_p]_0$}} \\
\colhead{(mag)} & \colhead{(mag)} & \colhead{(mag)} & \colhead{(mag)}
}
\startdata
5.0 &-0.3 &4.3      &0.55\\
5.0 &0.1  &4.0      &0.67\\
-6.0&0.2  &2.0      &0.72\\
-8.5&0.6  &0.0      &0.83\\
 -4 &3.0  &-0.9      &0.95\\
 10 &3.0  &-1.0      &1.1\\
 10 &-0.3 &8.0     &2.0\\
 $\cdots$&$\cdots$&15.0 &2.0\\
 $\cdots$&$\cdots$&16.0 &1.3\\
 $\cdots$&$\cdots$&13.0 &0.55\\
\enddata
\end{deluxetable}

\subsection{Output clean YSOC catalog}\label{sec:cleanyso}

\begin{figure*}
    \centering
    \includegraphics[width=1.0\linewidth]{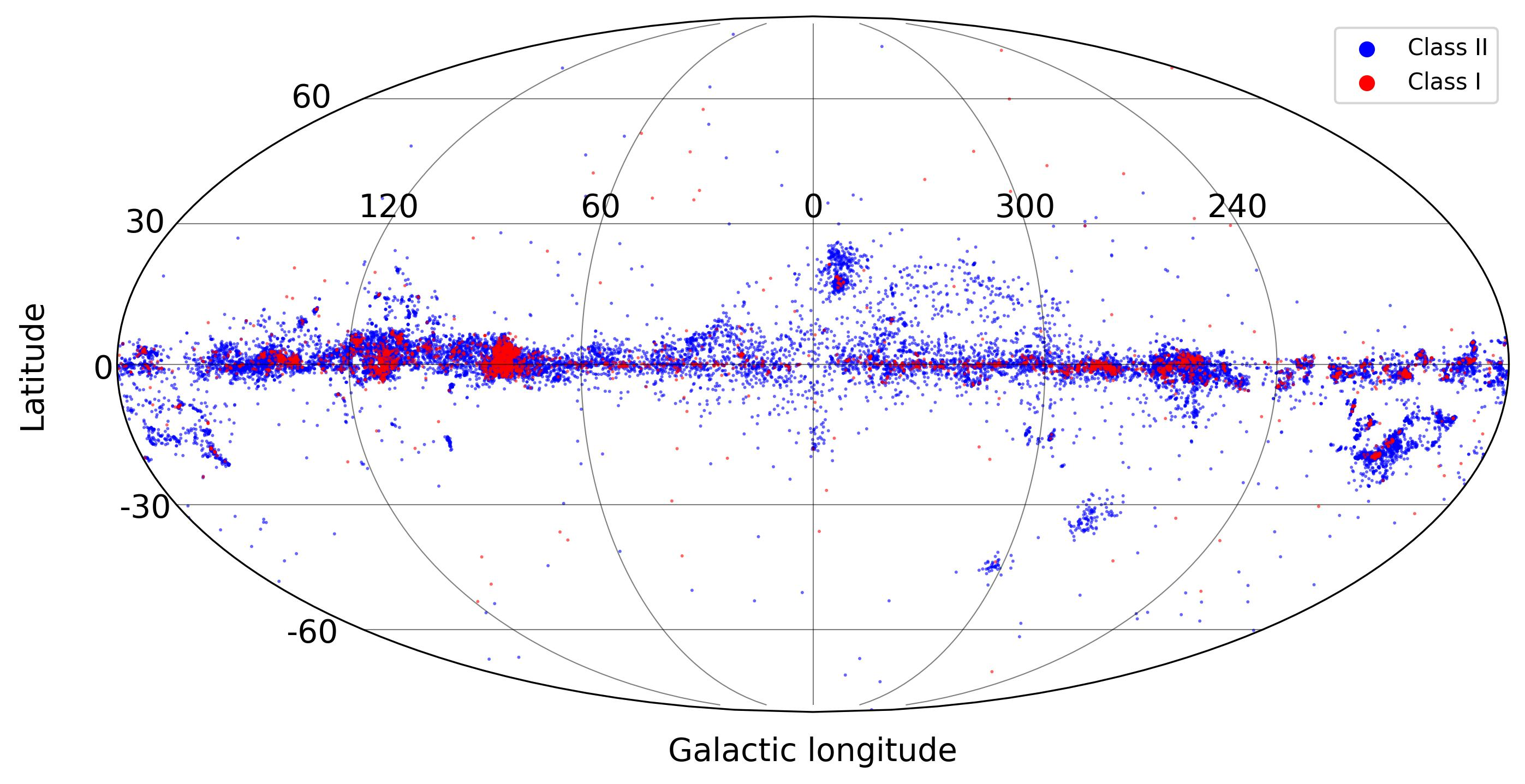}
    \caption{The spatial distribution of the clean YSOC sample in the mollweide projection. The Class I and Class II candidates are labeled with red and blue filled circles, repectively.}
    \label{fig:ysocleandistribution}
\end{figure*}

The final clean YSOC catalog has \mz{24\,883 }
YSOCs. Table~\ref{tab:ysoclean} shows the entries of the catalog, including the {\it Gaia}, 2MASS, and WISE photometries, {\it Gaia} parallaxes and proper motions, \mz{geometric distances from \citet{gaiadist2021}}, and foreground extinctions and spectral indices estimated in Sect.~\ref{sec:ysoc-deredden}. I also calculate the Galactocentric coordinates of YSOCs with the solar motion parameters suggested by \citet{reid2019}, assuming that the sun is located on the x axis of the right-handed system. Figure~\ref{fig:ysocleandistribution} shows the spatial distribution of this clean YSOC sample. For convenience, I simply use ``YSOC catalog" to refer this ``clean YSOC catalog" in the subsequent context.

\mz{As mentioned in Sect.~\ref{sec:contamex}, my method finally results in a relatively clean but less complete YSO sample. Many luminous PMS stars have been removed from my YSOC catalog and there are still MS residuals in the catalog. I compare my YSOC catalog with the YSOs in Orion A \citep{ysooriona2019}, 
which can be used to \mztwo{infer} the completeness and contamination level of my YSOC catalog.}

\mz{\citet{ysooriona2019} compiled a list of YSOs in Orion A molecular cloud by combining the deep near-infrared (NIR) VISTA survey data \citep[VISION,][]{vision2016} and archival mid-infrared (MIR) to far-infrared (FIR) data such as {\it Spitzer}, {\it Herschel}, and WISE. They carefully revisited the known YSOs in literature with the aim to evaluate false positives, and then added new YSOs that were obtained with NIR and MIR color criteria. \citet{ysooriona2019} finally obtained 2\,980 YSOs in Orion A. I note that their sample was spatially biased due to the different coverage of infrared surveys. To get a YSO sample with roughly uniform completeness, I extract 2849 YSOs inside the {\it Spitzer}/IRAC data coverage. As a comparison, there are 718 sources of my YSOC catalog located in the same {\it Spitzer}/IRAC coverage. I crossmatch these 718 YSOCs with \citet{ysooriona2019}'s 2849 YSOs and find that there are 482 sources in common. 

Assuming that all YSOs presented by \citet{ysooriona2019} are bona-fide young stars, the contamination fraction of my YSOC catalog is about 30\% in Orion A. \mztwo{This percentage (30\%) should be considered only as a rough estimate of the contamination fraction in the entire YSOC catalog, as it does not account for variations in distance and star formation environments across different molecular clouds.}
\citet{ysooriona2019} also estimated the completeness of their YSO sample in the {\it Spitzer}/IRAC coverage to be about 49\%. Therefore, the completeness of my YSOC catalog is $\sim$10\% in Orion A, which infers that the completeness of the whole YSOC catalog could be $<$10\% considering the distance of Orion A \citep[$\sim$430\,pc,][]{zucker2019}.}


My YSOC catalog has potential for use in future follow-up observations and for statistical studies such as investigating star formation in the solar neighborhood. However, there are \mztwo{three} caveats that need to be taken into account when using this catalog. First, more than 80\% of the sources in my YSOC catalog come from \citet{ysowise2019}'s {\it Gaia}-AllWISE YSOC sample, which was only identified in areas of the sky above a certain dust opacity threshold based on the Planck dust map. Therefore, both the {\it Gaia}-AllWISE YSOC sample and my YSOC catalog suffer from spatial bias. Second, my method results in the loss of true luminous YSOs, implying that my YSOC catalog is biased towards low-mass young stars. \mztwo{Third, it is important to note that the sources in my YSOC catalog are YSO candidates rather than confirmed young stars. This means that there is a potential for high contamination (e.g., $\sim$30\% in Orion A). However, without additional spectroscopic information, it is difficult to isolate bona-fide YSOs. Consequently, any statistical analysis based on my YSOC catalog, such as cloud distance estimation (see Sect.~\ref{sec:results}), is inevitably affected by the potentially high level of contamination.}

\startlongtable
\begin{deluxetable*}{lll}
\tablecaption{Entries of the clean YSOC catalog\label{tab:ysoclean}}
\tablewidth{0pt}
\tablehead{
\colhead{Entry} & \colhead{Units}  & \colhead{Description}
}
\startdata
AllWISE & $\ldots$ & AllWISE catalog name \\
RAJ2000 & deg & Right ascension (J2000) \\
DEJ2000 & deg & Declination (J2000) \\
Glon & deg & Galactic longitude \\
Glat & deg & Galactic latitude \\
$X$ & kpc & Galactocentric $x$ position component\\
$Y$ & kpc & Galactocentric $y$ position component\\
$Z$ & kpc & Galactocentric $z$ position component\\
$J$mag & mag & 2MASS $J$ band magnitude\\
e\_$J$mag & mag & Uncertainty of $J$ magnitude\\
$H$mag & mag & 2MASS $H$ band magnitude\\
e\_$H$mag & mag & Uncertainty of $H$ magnitude\\
$K$mag & mag & 2MASS $K_s$ band magnitude\\
e\_$K$mag & mag & Uncertainty of $K_s$ magnitude\\
$W1$mag & mag & WISE $W1$ band magnitude\\
e\_$W1$mag & mag & Uncertainty of $W1$ magnitude\\
$W2$mag & mag & WISE $W2$ band magnitude\\
e\_$W2$mag & mag & Uncertainty of $W2$ magnitude\\
$W3$mag & mag & WISE $W3$ band magnitude\\
e\_$W3$mag & mag & Uncertainty of $W3$ magnitude\\
$W4$mag & mag & WISE $W4$ band magnitude\\
e\_$W4$mag & mag & Uncertainty of $W4$ magnitude\\
Ref & $\ldots$ & References\\
\mz{GaiaDR3\_source\_id} & $\ldots$ & Unique source identifier in {\it Gaia} \mz{DR3}\\
\mz{fidelity\_v2} & $\ldots$ & \mz{Astrometric fidelities}\\
\mz{norm\_dg} & $\ldots$ & \mz{Diagnostics of contamination from neighbors}\\
Plx & mas & Column $\mathtt{parallax}$ in {\it Gaia} \mz{DR3}\\
e\_Plx & mas & Column $\mathtt{parallax\_error}$ in {\it Gaia} \mz{DR3}\\
$G$mag & mag & Column $\mathtt{phot\_g\_mean\_mag}$ in {\it Gaia} \mz{DR3}\\
e\_$G$mag & mag & \mz{Uncertainty of $G$mag}, \mzac{see Sect.~\ref{sec:gaiadr2cat}}\\
$Bp$mag & mag & Column $\mathtt{phot\_bp\_mean\_mag}$ in {\it Gaia} \mz{DR3}\\
e\_$Bp$mag & mag & \mz{Uncertainty of $Bp$mag}, \mzac{see Sect.~\ref{sec:gaiadr2cat}}\\
$Rp$mag & mag & Column $\mathtt{phot\_rp\_mean\_mag}$ in {\it Gaia} \mz{DR3}\\
e\_$Rp$mag & mag & \mz{Uncertainty of $Rp$mag}, \mzac{see Sect.~\ref{sec:gaiadr2cat}}\\
pmRA & mas\,yr$^{-1}$ & Column $\mathtt{pmra}$ in {\it Gaia} \mz{DR3}\\
e\_pmRA & mas\,yr$^{-1}$ & Column $\mathtt{pmra\_error}$ in {\it Gaia} \mz{DR3}\\
pmDE & mas\,yr$^{-1}$ & Column $\mathtt{pmdec}$ in {\it Gaia} \mz{DR3}\\
e\_pmDE & mas\,yr$^{-1}$ & Column $\mathtt{pmdec\_error}$ in {\it Gaia} \mz{DR3}\\
\mz{r\_med\_geo} & \mz{pc} & \mz{Median geometric distance} \\
\mz{r\_lo\_geo} & \mz{pc} & \mz{16th percentile of geometric distance}\\
\mz{r\_hi\_geo} & \mz{pc} & \mz{84th percentile of geometric distance}\\
\mz{flag} & $\ldots$ & \mz{Flag of geometric distance}\\
\mz{zpt} & \mz{mas} & \mz{Zero point of parallax bias}\\
\mz{$A_G$\_final} & mag & Foreground extinction\\
\mz{e\_$A_G$\_final} & mag & Uncertainty of extinction\\
alpha & $\ldots$ & Observed spectral index\\
\mz{e\_alpha} & $\ldots$ & \mz{Uncertainty of observed spectral index}\\
alphac & $\ldots$ & De-reddened spectral index\\
\mz{e\_alphac} & $\ldots$ & \mz{Uncertainty of de-reddened spectral index}\\
\enddata
\tablecomments{The full catalog can be derived online in the China–VO PaperData repository: \mzac{doi:} \href{https://doi.org/10.12149/101211}{\mzac{10.12149/101210}}. 
(This table is available in its entirety in machine-readable form)}
\end{deluxetable*}

\section{Cloud distance estimation}\label{sec:results}
In Section \ref{sec:ysocat}, I presented an all-sky YSOC catalog. In the subsequent sections, I utilized this catalog to estimate the distances of several tens of nearby molecular clouds. In Section \ref{sect:sample}, I outline the sample selection process for the nearby molecular clouds and the methodology used to determine their boundaries. Section \ref{sect:ysoinclouds} explains how YSOCs were isolated within the local clouds, and Section \ref{sect:clouddist} describes the method used to estimate cloud distances. Finally, in Section \ref{sec:clouddistcat}, I present the catalog of cloud distances.

\subsection{Sample of local clouds}\label{sect:sample}

The sample of the local molecular clouds is constructed with the cloud catalogs released by \citet{zucker2019,zucker2020} and \citet{spilker2021}. \citet{zucker2019} obtained the accurate distances of 27 nearby molecular clouds that were inferred with the distance and extinction of stars along the sightlines towards clouds based on the stellar photometric catalog and the {\it Gaia} DR2 parallax measurements. \citet{zucker2020} applied the method suggested by \citet{zucker2019} to the star forming regions described in the Star Formation Handbook \citep{sfbook_north,sfbook_south} and obtained the accurate distances to $\sim$60 local star-forming regions. \citet{spilker2021} compiled a catalog of nearby molecular clouds, including 72 clouds, and analyzed their column density probability distributions. I combine these three catalogs and construct a sample of local molecular clouds, including \mz{about a hundred} clouds. 



\begin{figure*}
    \centering
    \includegraphics[width=1.0\linewidth]{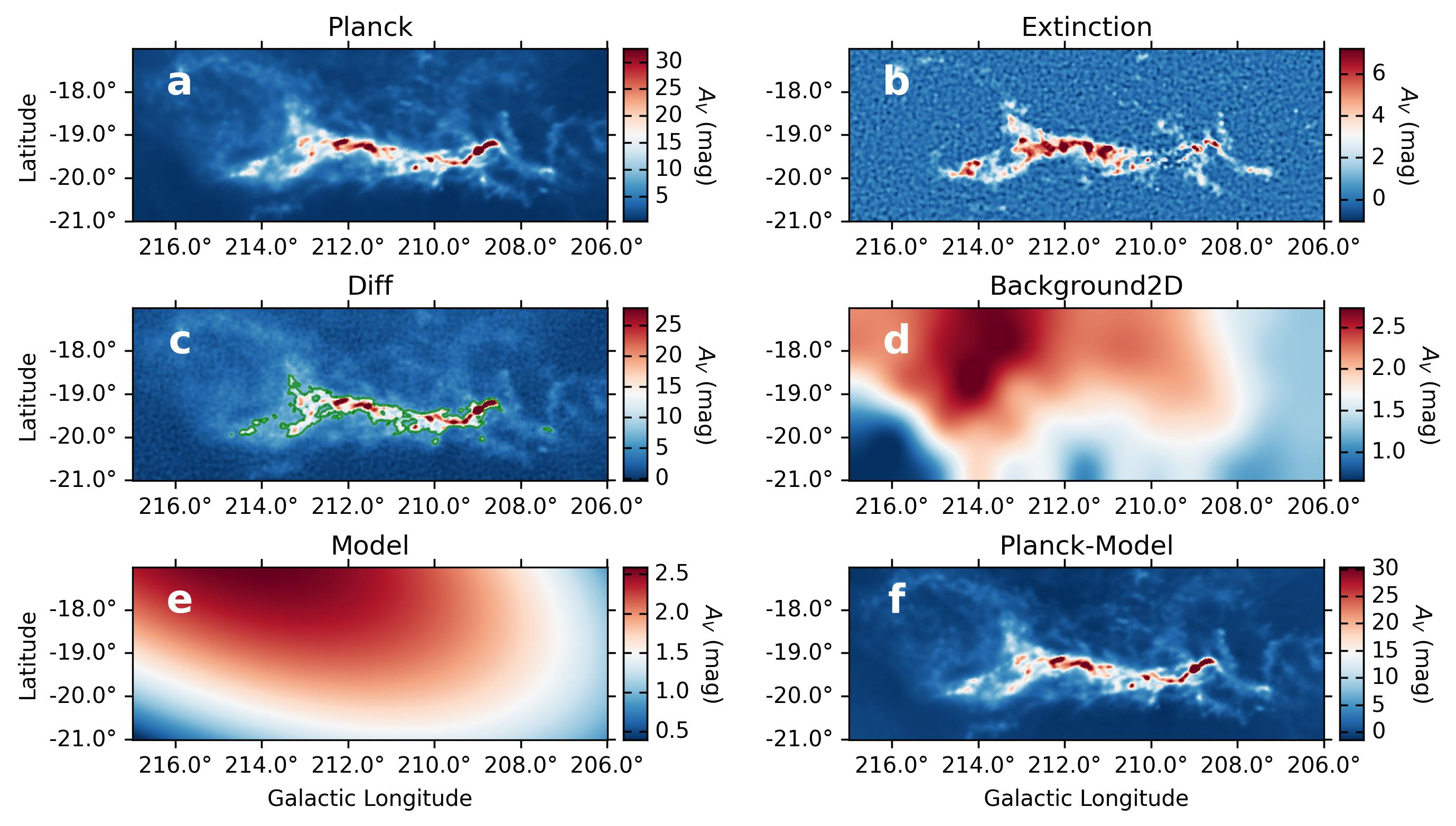}
    \caption{Method used to obtain the background-subtracted Planck dust map for the Orion A molecular cloud. (a): The Planck dust map; (b): the extinction map from \citet{dobashi2011}; (c): difference map between the Planck dust map and extinction map; (d): 2D background of the difference map calculated with a mode estimator using the Source Extractor \citep{sex1996} algorithm after masking 5\% pixels with highest $A_V$ values; (e): the background model obtained by fitting the 2D background with a 2D polynomial function; (f): the modeled-background-subtracted Planck dust map. The green contours in panel c mark the 95\% percentile of $A_{V}$ values in the difference map.}
    \label{fig:extmap_oriona}
\end{figure*}

I use the Planck dust map (as described in Sect.\ref{sec:plankdustmap}) and the extinction map\footnote{\url{http://darkclouds.u-gakugei.ac.jp/}} by \citet{dobashi2011} to define the boundaries of the local clouds in my sample. To illustrate, Fig.\ref{fig:extmap_oriona}a displays the Planck dust map for the Orion A molecular cloud, which provides the total column density along lines of sight. However, because of contamination from the diffuse dust component, I cannot define cloud boundaries directly with the Planck dust map. In contrast, \citet{dobashi2011} produced an all-sky extinction map based on the 2MASS \citep{2mass} point source catalog that eliminates extinction from the diffuse dust component \citep{dobashi2013}. Thus, their extinction map can trace the cloud column density. However, due to the limited sensitivity of the 2MASS survey and the technique used for extinction mapping \citep{dobashi2008,kainulainen2011}, it cannot effectively trace the dense structures in molecular clouds. Figure~\ref{fig:extmap_oriona}b shows the \citet{dobashi2011} extinction map for the Orion A cloud, which reveals that the dense integral-shaped filament (ISF) of Orion A, clearly visible in the Planck dust map, corresponds to an abnormal low extinction region in the extinction map. As a result, using the extinction map to define cloud boundaries could lead to missing the dense regions that are likely closely associated with the YSOs \citep{gao2004,lada2010,mypub2019}.

To delineate the local clouds reasonably, I devise a method that combines the Planck dust maps with \citet{dobashi2011}'s extinction maps. This involves subtracting the diffuse dust component from the Planck dust map using the extinction map as a reference. Fig.~\ref{fig:extmap_oriona} demonstrates this process for the Orion A cloud. First, I obtain a difference map by subtracting the extinction map from the Planck dust map (Fig.~\ref{fig:extmap_oriona}c). The difference map highlights the dense regions and the diffuse dust component. Second, I estimate the two-dimensional (2D) background of the difference map using a mode estimator \citep[implemented in Source Extractor,][]{sex1996} and masking the 5\% pixels with the highest $A_V$ values (Fig.~\ref{fig:extmap_oriona}d). Third, I fit the 2D background map with a 2D polynomial function to obtain a background model (Fig.~\ref{fig:extmap_oriona}e). Finally, I subtract the background model from the Planck dust map to produce the result shown in Fig.~\ref{fig:extmap_oriona}f. 

I generate background-subtracted Planck dust maps for all the local clouds in my sample. To define the boundaries of these clouds, I use the extinction contour level of $A_V=$~2 mag, which was recommended by both \citet{heiderman2010} and \citet{evans2014}. 

\subsection{YSOCs likely associated with the local clouds}\label{sect:ysoinclouds}

In Section~\ref{sec:cleanyso}, I estimated that the fraction of contamination, such as MS stars, in my YSOC catalog could be as high as 30\%. In this section, I aim to remove this contamination from the YSOC catalog for each local cloud using additional astrometric information from {\it Gaia} DR3. The fundamental assumption is that YSOs in the same local cloud should have comparable parallaxes and proper motions. 
In the subsequent context, I use the astrometric notation $\alpha$ and $\delta$ for Right ascension and Declination, $\varpi$ for parallax in units of mas, $\mu_{\alpha}\cos \delta$ and $\mu_{\delta}$ for proper motions in units of mas yr$^{-1}$. 
\mz{I limit my selection to the YSOCs with {\it Gaia} DR3 parallax and proper motion measurements. I also apply a quality cut of $\varpi>0$ as suggested by \citet{prisinzano2022}. Considering that I only focus on the YSOs in the nearby molecular clouds, this choice does not introduce any bias. Finally I obtained 23\,379 YSOCs from my YSOC catalog as the input sample.}

\begin{figure*}
\includegraphics[width=1.0\linewidth]{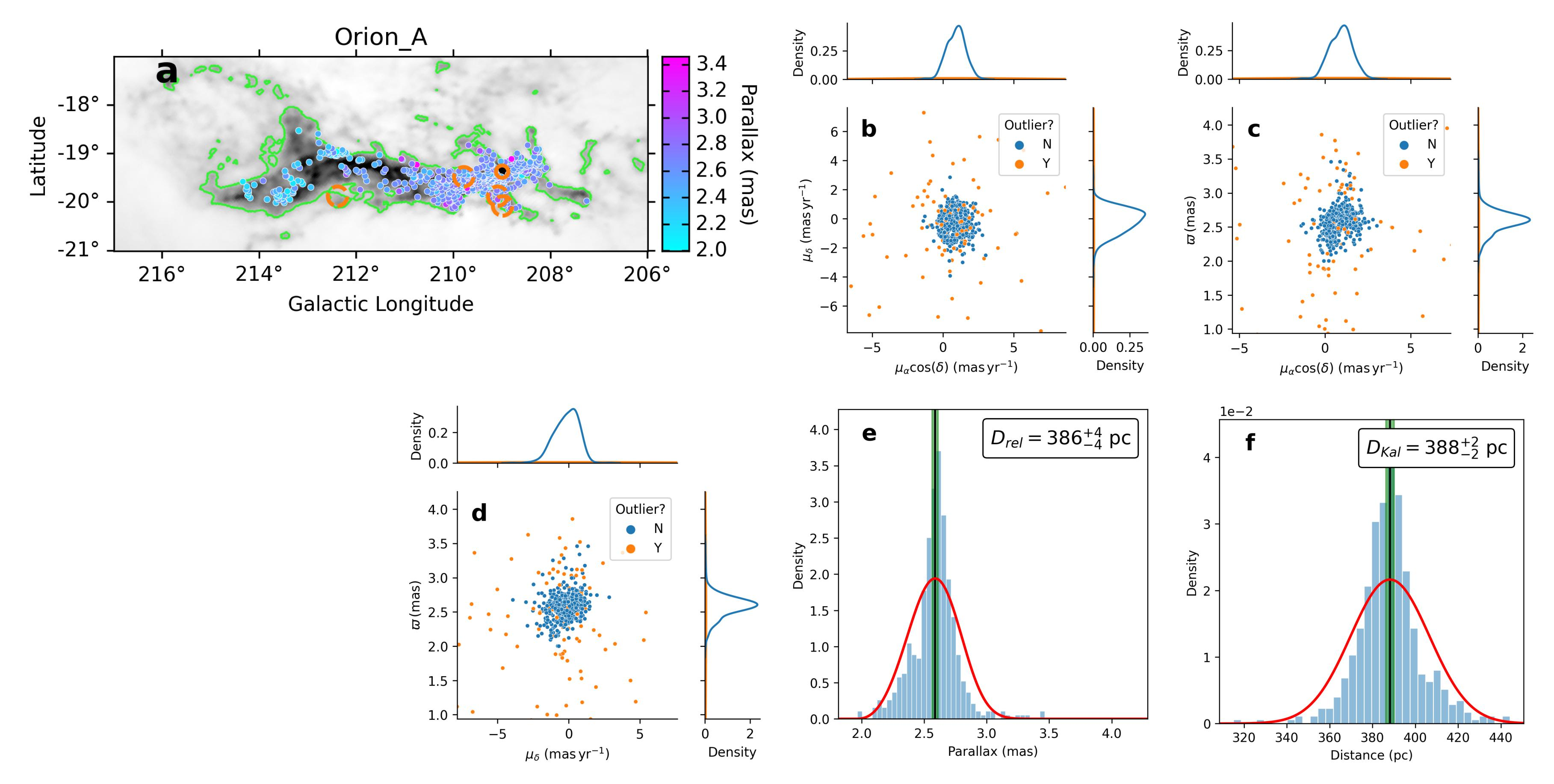}
\caption{\mz{Method used to isolate the YSOCs in the Orion A molecular cloud and estimate the distance of Orion A. 
        (a): The background-subtracted Planck dust map, overlaid with the YSOCs that are likely to be associated 
        with the Orion A cloud. The green contour label the level of $A_V=$~2 mag. The YSOCs are marked with 
        filled circles, color-coded by their Gaia parallax. The orange dashed circles mark the sightline beams 
        towards which \citet{zucker2020} 
 obtained the Bayesian distances while the orange solid circles label 
        the stellar clusters with distances by \citet{kuhn2019} 
; (b): proper motion distribution of YSOCs in the 
        region of Orion A with $A_V>$~2 mag. The side panels show the KDEs of proper motions 
        in R.A. (top) and decl. (right). The orange dots mark the outliers identified with DBSCAN technique 
        \citep{dbscan1996}; (\mztwo{c and d}): proper motion versus parallax of YSOCs in Orion A. The side panels show the KDEs of proper motions 
        in R.A. \mztwo{or decl.} (top) and parallax (right). The orange dots mark the outliers identified with DBSCAN algorithm; 
        (\mztwo{e}): parallax distribution of YSOCs in Orion A after outlier removal. 
        The red curve shows the PDF obtained by fitting the YSOC parallaxes with the Weibull model. 
        The vertical solid line mark the 
        mode parallax to the Orion A while the green shaded area label the uncertainty of the parallax. The
        corresponding distance of the mode parallax is marked on the panel. (\mztwo{f}): distance distribution of YSOCs
        obtained with Kalkayotl in Orion A after outlier removal. The red curve shows the PDF obtained by fitting
        the YSOC parallaxes with the Kalkayotl program. The vertical solid line mark the distance to the Orion A
        while the green shaded area label the uncertainty of the distance. The value of the distance is also 
        marked on the panel. The complete figure set (63 images) is available in the online journal.} 
\label{fig:oriona_dist}}
\end{figure*}

\mztwo{I use the coordinates, parallaxes, and proper motions of YSOCs to filter out contamination from the input sample in each local cloud. To illustrate, I show the detailed filtering process for the Orion A molecular cloud in Fig.~\ref{fig:oriona_dist}. First, I select YSOCs that are inside the boundary of each cloud, resulting in approximately 4800 YSOCs in about 70 local clouds, noting that around 30 nearby clouds in my cloud sample (as described in Sect.~\ref{sect:sample}) have no YSOCs. These clouds are mostly quiescent and without active star formation, such as Pegasus, Aquila South, and Draco. Some nearby clouds, such as Chamaeleon, Lupus, and Cepheus, have different levels of star formation activity in different parts. For these clouds, I only extract sub-clouds with YSOCs, such as Cham I, Lupus I, and Cepheus-L1251.}  

\mztwo{Second, I use DBSCAN \citep{dbscan1996} as implemented in scikit-learn \citep{scikit-learn} to remove outliers in the 3D parameter space ($\varpi$, $\mu_{\alpha}\cos \delta$, $\mu_{\delta}$) of YSOCs in each local cloud. DBSCAN identifies core samples with more than $minPts$ points within a radius $\epsilon$ of a given point $\vec{p}$, and constructs clusters with sets of core samples. Points that are not included in any clusters are treated as outliers. The values of $\epsilon$ and $minPts$ are critical in identifying clusters.}


\mz{The three parameters ($\varpi$, $\mu_{\alpha}\cos \delta$, $\mu_{\delta}$) of YSOCs in each local cloud are first re-scaled using the tool of scikit-learn, $\mathtt{RobustScaler}$, which is based on the statistics robust to outliers. The value of $minPts$ defines the minimal number of a cluster. I adopt $minPts=$~6 that is twice of the dimensions of parameter space \citep{Sander1998DensityBasedCI}. The value of $\epsilon$ is determined using the $k$-distance method \citep{Rahmah2016}. The $k$th nearest neighbor distance ($k$-distance) can be calculated for each point in ($\varpi$, $\mu_{\alpha}\cos \delta$, $\mu_{\delta}$) space. If plotting these $k$-distances in ascending order, a sharp change of slope, i.e., the knee point, can be found along the $k$-distance curve. I use the Python code $\mathtt{kneed}$\footnote{\url{https://github.com/arvkevi/kneed}} \citep{kneed} to detect the knee point automatically and then this knee point is adopted as the optimal value of $\epsilon$. \citet{Sander1998DensityBasedCI} found that the $k$ value does not significantly affect the DBSCAN results and thus is not very crucial for the algorithm. I tried several different values of $k \in [1,6]$ and found that $k=1$ can remove the outliers more efficiently. Therefore, I finally use $k=1$ to calculate the optimal value of $\epsilon$.}

\mz{The DBSCAN algorithm itself does not consider the uncertainties of the parameter. However, the uncertainties of $\varpi$, $\mu_{\alpha}\cos \delta$ and $\mu_{\delta}$ in my YSOC catalog could be relatively large given that I do not perform any quality cuts on their uncertainties. To include the effect of uncertainties, I use a Monte Carlo method to remove the outliers with DBSCAN. Specifically, I generate a random set of ($\varpi$, $\mu_{\alpha}\cos \delta$, $\mu_{\delta}$) in each local cloud by assuming a gaussian error distribution. The outliers can be marked after running DBSCAN. Repeat the above process 1000 times and then a outlier probability ($P_{\mathrm{outlier}}$) can be obtained for each YSOC. I calculate the median of $P_{\mathrm{outlier}}$ of YSOCs ($P_{\mathrm{outlier,med}}$) in each local cloud and require $P_{\mathrm{outlier}}<P_{\mathrm{outlier,med}}$ to filter out the contamination of YSOCs.} 
\mz{Figure~\ref{fig:oriona_dist}\mztwo{b, c, and d} show the parallax and proper motion distributions of YSOCs in the region of Orion A with $A_V>$~2 mag and the identified outliers are also marked.} 


Finally I obtain \mztwo{3\,144 }
YSOCs that are likely to be associated with \mz{63} nearby molecular clouds. 
Table~\ref{tab:ysocinclouds} lists their \mz{information, including AllWISE names, parent cloud names, distances obtained with {\it Kalkayotl} (see Sect.~\ref{sect:clouddist}), and the heliocentric positions}.  Further YSOC information such as the photometry and {\it Gaia} \mz{DR3} parameters can be obtained by cross-matching with Table~\ref{tab:ysoclean} using the AllWISE name.

\begin{deluxetable*}{cccccc}
\tablecaption{{YSOCs likely associated with the local clouds}\label{tab:ysocinclouds}}
\tablehead{
\colhead{AllWISE} & \colhead{cloud} &
\colhead{$X_{\mathrm{H}}$\tablenotemark{a}} & \colhead{$Y_{\mathrm{H}}$\tablenotemark{a}} & \colhead{$Z_{\mathrm{H}}$\tablenotemark{a}} &
\colhead{$D_{\mathrm{Kal}}$\tablenotemark{b}}\\
\colhead{}&\colhead{}&\colhead{(pc)}&\colhead{(pc)}&\colhead{(pc)}&\colhead{(pc)}
}
\startdata
J000017.17+673045.8 & Cep\_OB4 & -505 &  948 &  96 &   $1078_{-61}^{+65}$ \\
J000040.95+664407.6 & Cep\_OB4 & -535 & 1009 &  87 & $1145_{-247}^{+310}$ \\
J000100.32+671415.5 & Cep\_OB4 & -703 & 1318 & 127 & $1499_{-225}^{+283}$ \\
J000112.52+673732.4 & Cep\_OB4 & -640 & 1196 & 124 & $1362_{-268}^{+319}$ \\
J000129.32+665426.9 & Cep\_OB4 & -693 & 1301 & 116 & $1479_{-205}^{+252}$ \\
J000145.28+664748.7 & Cep\_OB4 & -537 & 1007 &  88 & $1144_{-280}^{+332}$ \\
J000148.35+672728.1 & Cep\_OB4 & -492 &  918 &  92 &   $1046_{-56}^{+60}$ \\
J000152.18+672845.7 & Cep\_OB4 & -474 &  884 &  89 & $1006_{-224}^{+270}$ \\
J000200.37+672356.9 & Cep\_OB4 & -372 &  693 &  69 &    $790_{-55}^{+62}$ \\
J000207.32+672259.6 & Cep\_OB4 & -632 & 1178 & 116 & $1342_{-280}^{+326}$ \\
\enddata
\tablenotetext{a}{\mz{The heliocentric positions of the YSOCs. Here I define a heliocentric coordinate system ($X_{\mathrm{H}}, Y_{\mathrm{H}}, Z_{\mathrm{H}}$) with the Sun at the origin. The $X_{\mathrm{H}}$ axis points from the Sun to the Galactic center and the $Y_{\mathrm{H}}$ axis points roughly towards the Galactic longitude of 90\degr. The $Z_{\mathrm{H}}$ axis is orthogonal to the Galactic plane, pointing to the North Galactic Pole.}}
\tablenotetext{b}{\mz{The distances estimated with {\it Kalkayotl} program. See text for details.}}
\tablecomments{The machine-readable table can be accessed online in the China–VO PaperData repository: \mzac{doi:} \href{https://doi.org/10.12149/101211}{\mzac{10.12149/101210}}. 
A portion is shown here for guidance regarding its form and content.}
\end{deluxetable*}

\subsection{Estimation of distances to local clouds}\label{sect:clouddist}

The YSOCs listed in Table~\ref{tab:ysocinclouds} are the youngest optically visible sources in the nearby molecular clouds (see Sect.~\ref{sect:ysoinclouds}). Therefore, they are good proxies of the cloud distances. 

I use {\it Kalkayotl}\footnote{\url{https://github.com/olivares-j/kalkayotl}} to estimate the cloud distances. {\it Kalkayotl} is a free and open code developed by \citet{kalkayotl2020}. It is specifically designed to estimate cluster parameters such as size and distance, as well as the distances to individual members based on their {\it Gaia} parallax measurements. {\it Kalkayotl} employs a Bayesian hierarchical model to obtain the posterior distributions of distances for both the cluster and its members. The code utilizes distance prior families that are optimized for clusters and accounts for the spatial correlations of parallaxes. \citet{kalkayotl2020} have demonstrated that {\it Kalkayotl} can provide high credibility distance estimates for stellar clusters located within 5 kpc and with a size of $<$1 kpc.

{\it Kalkayotl} needs an initial guess of the cluster distance to construct the prior distribution. Therefore, I first derive a median distance for each local cloud by modeling the parallax distribution of YSOCs in each cloud. Specifically, I make the assumption that the parallax distribution of YSOCs in a given local cloud, as shown in Figure~\ref{fig:oriona_dist}d, is drawn from a Weibull probability density function (PDF):
\begin{eqnarray}
f(x) = \frac{\beta}{\eta}\left( \frac{x-\gamma}{\eta} \right)^{\beta-1} e^{-\left( \frac{x-\gamma}{\eta} \right)^{\beta}},
\end{eqnarray}
where $\eta$, $\beta$, and $\gamma$ are scale, shape, and location parameters, respectively. The Weibull distribution is a highly adaptable distribution that can replicate a diverse range of distributions through adjusting the shape parameter value. This distribution has the capability to model a broad spectrum of distributions and, as a result, has been extensively utilized in data analysis. I fit the YSOC \mz{parallax} distribution with Weibull PDF using the maximum likelihood estimation technique as implemented in the \textit{reliability}\footnote{\url{https://github.com/MatthewReid854/reliability}} python library \citep{reliability}. In specific, 
\mz{the parallax distribution of YSOCs is fitted twice} in each cloud: first with a three-parameter Weibull PDF; and second with a two-parameter Weibull PDF that forces $\gamma=$~0. I always adopt the fitting result with lower Bayesian information criterion (BIC). I also require $\beta>$~1 in the whole fitting process to avoid infinite probability.
The red curve in Fig.~\ref{fig:oriona_dist}e shows the Weibull PDF defined by the fitting parameters. 
\mz{Then I} use the Monte Carlo method to estimate the cloud \mz{median parallax} and their uncertainties. In each local cloud, I generate 10\,000 random Weibull PDFs assuming a gaussian error for $\eta$, $\beta$, and $\gamma$. Then the mode value is calculated for each random Weibull PDF. The cloud \mz{median parallax ($\varpi_{rel}$)} and its uncertainty are adopted as the median and standard deviation of 10\,000 mode values. \mz{I adopt the inversion of the cloud median parallax as the initial guess of the cloud distance, i.e., $D_{rel}$}.%

\mz{Using the implemented Gaussian prior model with a mean distance of $D_{rel}$, {\it Kalkayotl} (version 1.1) reports the samples of the posterior distribution of distance for each local cloud. The final cloud distance ($D_{\mathrm{Kal}}$) and associated uncertainty are calculated based on the median and central 68\% quantiles of samples. {\it Kalkayotl} also provides the distance estimate for each YSOC in local clouds. Figure~\ref{fig:oriona_dist}f as an example shows the {\it Kalkayotl} distance distribution of YSOCs in Orion A.} 

\subsection{Catalog of distances to local clouds}\label{sec:clouddistcat}
\begin{figure*}
    \centering
    \includegraphics[width=1.0\linewidth]{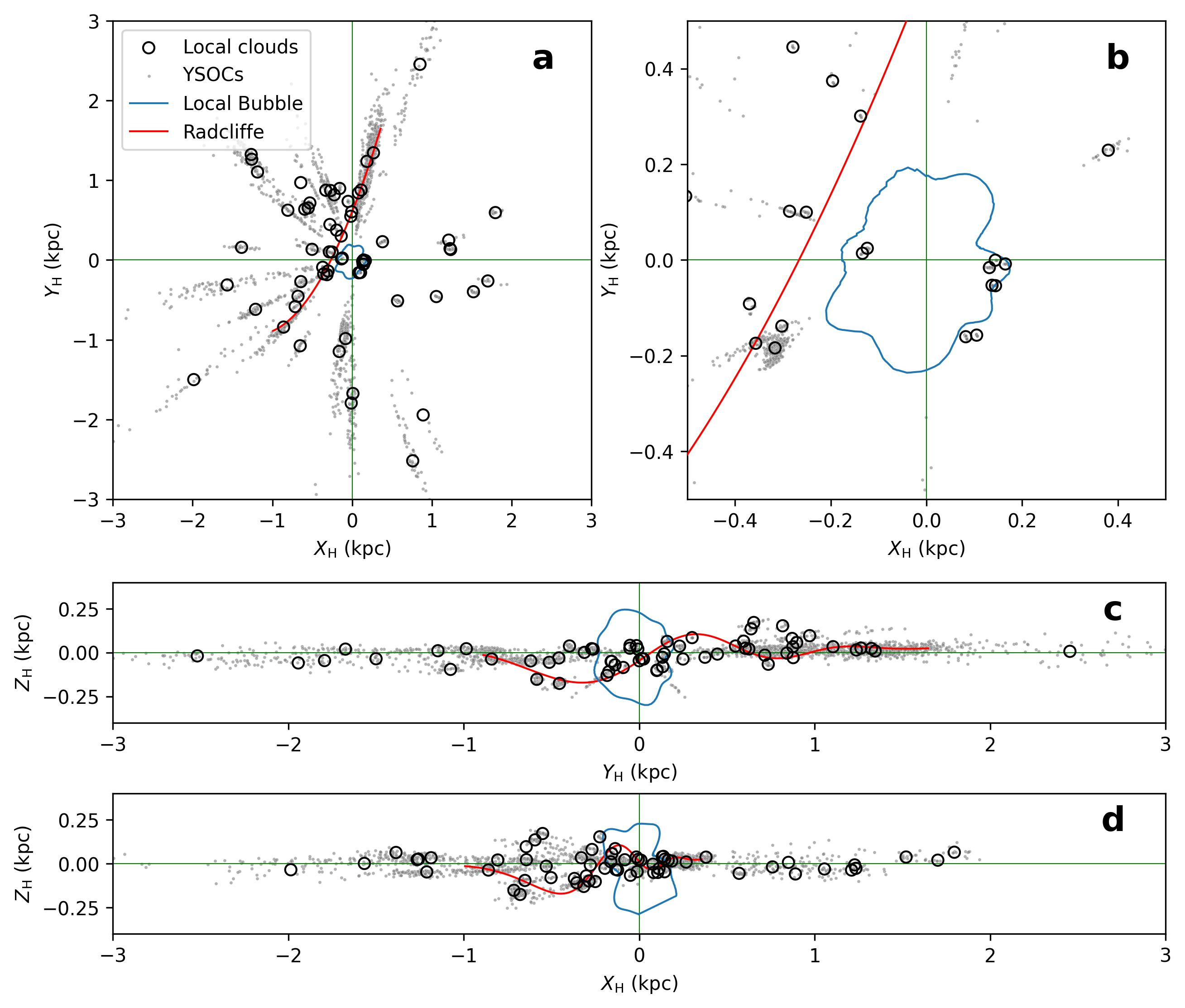}
    \caption{3D distribution of \mz{63} local clouds. The Sun is at ($X_{\mathrm{H}}$, $Y_{\mathrm{H}}$, $Z_{\mathrm{H}}$) $=$ (0, 0, 0) position that is marked with the intersection of two green lines. The panel a shows the top-down view while panels c and d show the side views. The panel b is the zoom-in view of the central region within 500 pc of the Sun. The YSOCs in the local clouds are marked with gray dots and the local clouds are labeled with black circles. The light blue lines show the inner surface of the Local Bubble shell \citep{pelgrims2020,zuckernature2022} and the red curve represents the model of Radcliffe Wave \citep{radcliffe2020,zuckernature2022}.}
    \label{fig:cloud_distribution}
\end{figure*}

The obtained distances to \mz{63} local clouds are given in Table~\ref{tab:clouddist}. The solid vertical line in Fig.~\ref{fig:oriona_dist}\mz{f} marks the distance of the Orion A while the green shaded area labels the uncertainty of the distance. Figure~\ref{fig:cloud_distribution} shows the 3D distribution of these \mz{63} local clouds. I also show the inner surface of the Local Bubble shell modeled by \citet{pelgrims2020} and the Radcliffe Wave identified by \citet{radcliffe2020} in Fig~\ref{fig:cloud_distribution}. It seems that there is condensation of molecular clouds along the Radcliffe Wave, which indicates that my distance catalog can also trace the Radcliffe Wave. Additionally, it is apparent that local clouds within a distance of about 200 pc from the Sun lie on the surface of the Local Bubble, as suggested by \citet{zuckernature2022}. In their analysis of the 3D spatial distribution and kinematics of dense gas and young stars in the solar neighborhood, \citet{zuckernature2022} propose that the expansion of the Local Bubble has caused the surrounding interstellar medium to be swept up into an extended shell, which fragmented and collapsed to form the local clouds. My distance estimates for these local clouds further support this view.

\begin{longrotatetable}
\begin{deluxetable*}{lDDDDDcDDDDD}
\tablecaption{Distances to \mz{63} nearby molecular clouds\label{tab:clouddist}}
\tabletypesize{\scriptsize}
\tablehead{
\colhead{Cloud} & 
\multicolumn2c{Glon} & 
\multicolumn2c{Glat} & \multicolumn2c{$X_{\mathrm{H}}$} & 
\multicolumn2c{$Y_{\mathrm{H}}$} & \multicolumn2c{$Z_{\mathrm{H}}$} & 
\colhead{$N_{\mathrm{YSOC}}$} & \multicolumn2c{$\mu_{\alpha}\cos (\delta)$} & 
\multicolumn2c{$\mu_{\delta}$} & \multicolumn2c{$\varpi_{rel}$} & 
\multicolumn2c{$D_{rel}$} & 
\multicolumn2c{$D_{\mathrm{Kal}}$}\\ 
\colhead{} & 
\multicolumn2c{($\degr$)} & \multicolumn2c{($\degr$)} & \multicolumn2c{(pc)} & 
\multicolumn2c{(pc)} & 
\multicolumn2c{(pc)} & 
\colhead{} & 
\multicolumn2c{(mas yr$^{-1}$)} & \multicolumn2c{(mas yr$^{-1}$)} & 
\multicolumn2c{(mas)} & 
\multicolumn2c{(pc)}  & 
\multicolumn2c{(pc)}
} 
\decimalcolnumbers
\startdata
AFGL490 & 142.256 &   1.186 &  -807 &   625 &   21 &      66 &   0.052$\pm$1.634 &  -1.144$\pm$1.764 & 1.259$\pm$0.081 &    $794.4_{-48.1}^{+54.7}$ &   $1020.5_{-54.5}^{+57.6}$ \\
              Ara & 336.521 &  -1.504 &  1055 &  -458 &  -30 &       9 &   1.193$\pm$0.479 &  -4.402$\pm$0.241 & 0.904$\pm$0.029 &   $1105.9_{-34.9}^{+37.3}$ &   $1151.0_{-38.0}^{+38.3}$ \\
         CMa\_OB1 & 224.386 &  -1.722 &  -860 &  -842 &  -36 &     120 &  -3.307$\pm$1.324 &   1.009$\pm$0.726 & 0.946$\pm$0.025 &   $1056.8_{-27.3}^{+28.8}$ &   $1203.8_{-28.6}^{+29.1}$ \\
       California & 165.122 &  -8.688 &  -503 &   134 &  -80 &      20 &   2.763$\pm$0.612 &  -4.998$\pm$0.574 & 2.102$\pm$0.097 &    $475.7_{-21.0}^{+23.1}$ &    $526.5_{-24.2}^{+24.5}$ \\
           Carina & 286.771 &  -0.401 &   759 & -2519 &  -18 &      56 &  -6.624$\pm$1.032 &   2.646$\pm$0.579 & 0.448$\pm$0.023 & $2231.6_{-109.0}^{+120.8}$ & $2630.9_{-111.1}^{+124.1}$ \\
        Cartwheel & 345.332 &   1.387 &  1520 &  -398 &   38 &      13 &  -0.182$\pm$0.421 &  -1.324$\pm$0.685 & 0.662$\pm$0.028 &   $1511.1_{-60.9}^{+66.2}$ &   $1572.1_{-56.1}^{+57.8}$ \\
         Cep\_OB3 & 110.693 &   2.076 &  -330 &   875 &   34 &     118 &  -0.913$\pm$1.683 &  -2.420$\pm$1.170 & 1.267$\pm$0.043 &    $789.4_{-25.7}^{+27.5}$ &    $935.8_{-31.4}^{+31.4}$ \\
         Cep\_OB4 & 123.580 &   4.753 &  -645 &   971 &   97 &      27 &  -1.509$\pm$1.773 &  -1.320$\pm$1.127 & 1.034$\pm$0.087 &    $966.7_{-75.0}^{+88.7}$ &   $1169.5_{-79.4}^{+83.6}$ \\
    Cepheus-L1251 & 114.558 &  14.537 &  -138 &   301 &   86 &       8 &   6.446$\pm$0.216 &   0.953$\pm$0.394 & 2.927$\pm$0.063 &      $341.7_{-7.2}^{+7.5}$ &      $341.8_{-4.0}^{+3.9}$ \\
  Cepheus-NGC7129 & 105.397 &  10.250 &  -225 &   817 &  153 &       9 &  -1.763$\pm$0.274 &  -3.441$\pm$0.280 & 1.104$\pm$0.039 &    $905.6_{-30.8}^{+33.1}$ &    $861.2_{-21.3}^{+24.5}$ \\
            ChamI & 297.141 & -15.535 &    82 &  -160 &  -50 &      16 & -22.895$\pm$0.890 &   1.150$\pm$1.184 & 5.391$\pm$0.029 &      $185.5_{-1.0}^{+1.0}$ &      $187.1_{-1.0}^{+0.9}$ \\
           ChamII & 303.751 & -14.799 &   105 &  -157 &  -50 &       8 & -19.799$\pm$0.428 &  -7.630$\pm$0.484 & 5.125$\pm$0.039 &      $195.1_{-1.5}^{+1.5}$ &      $195.4_{-1.1}^{+1.2}$ \\
         Circinus & 317.972 &  -4.099 &   569 &  -513 &  -55 &      25 &  -3.196$\pm$0.455 &  -3.936$\pm$0.520 & 1.334$\pm$0.027 &    $749.7_{-14.7}^{+15.3}$ &    $767.9_{-13.8}^{+14.0}$ \\
       Crossbones & 219.266 &  -9.342 &  -714 &  -584 & -152 &      44 &  -3.342$\pm$0.287 &   0.600$\pm$0.509 & 1.113$\pm$0.026 &    $898.6_{-20.8}^{+21.8}$ &    $935.2_{-16.2}^{+17.3}$ \\
      Cygnus-West &  70.893 &   0.164 &   850 &  2454 &    7 &      89 &  -2.753$\pm$1.809 &  -5.189$\pm$1.543 & 0.407$\pm$0.106 & $2458.9_{-506.6}^{+861.6}$ & $2597.2_{-168.0}^{+187.0}$ \\
    CygnusX-North &  81.492 &   0.705 &   185 &  1237 &   15 &     252 &  -2.287$\pm$3.151 &  -3.722$\pm$3.196 & 0.697$\pm$0.049 &  $1434.8_{-95.0}^{+109.4}$ &   $1251.0_{-36.3}^{+38.1}$ \\
    CygnusX-South &  78.765 &   0.397 &   267 &  1344 &    9 &     249 &  -2.240$\pm$1.623 &  -4.143$\pm$1.703 & 0.772$\pm$0.037 &   $1295.2_{-59.0}^{+64.9}$ &   $1370.7_{-36.2}^{+37.7}$ \\
     G82.65-02.00 &  82.719 &  -1.833 &   112 &   877 &  -28 &      26 &  -1.358$\pm$2.885 &  -3.868$\pm$3.004 & 1.403$\pm$0.142 &    $712.8_{-65.3}^{+80.0}$ &    $884.7_{-58.7}^{+59.0}$ \\
         Gem\_OB1 & 191.301 &   0.051 & -1567 &  -313 &    1 &      90 &  -0.017$\pm$1.326 &  -1.683$\pm$3.048 & 0.585$\pm$0.096 & $1709.1_{-240.0}^{+333.7}$ &  $1597.8_{-97.6}^{+103.7}$ \\
           IC1396 &  99.912 &   3.614 &  -157 &   897 &   58 &       9 &  -2.012$\pm$1.139 &  -3.650$\pm$0.929 & 1.153$\pm$0.044 &    $867.1_{-32.2}^{+34.8}$ &    $912.6_{-38.1}^{+35.3}$ \\
           IC2944 & 294.597 &  -1.569 &   890 & -1944 &  -59 &      10 &  -6.214$\pm$0.265 &   0.789$\pm$0.304 & 0.615$\pm$0.071 & $1626.7_{-168.3}^{+212.2}$ & $2138.4_{-203.7}^{+210.6}$ \\
           IC5146 &  93.938 &  -5.119 &   -51 &   735 &  -66 &      12 &  -2.799$\pm$0.384 &  -2.727$\pm$0.320 & 1.333$\pm$0.022 &    $750.3_{-12.2}^{+12.6}$ &    $739.6_{-12.0}^{+12.2}$ \\
     L1003and1004 &  91.764 &   3.950 &   -17 &   548 &   38 &      10 &   1.153$\pm$1.359 &  -3.134$\pm$0.708 & 1.836$\pm$0.026 &      $544.5_{-7.7}^{+7.9}$ &      $549.9_{-7.4}^{+7.2}$ \\
            L1265 & 117.638 &  -3.547 &  -196 &   374 &  -26 &       8 &   8.208$\pm$1.452 &  -2.414$\pm$0.561 & 2.538$\pm$0.158 &    $394.1_{-23.1}^{+26.2}$ &    $423.5_{-14.3}^{+17.0}$ \\
            L1302 & 122.074 &  -0.886 &  -279 &   445 &   -8 &       6 &   3.522$\pm$0.181 &  -2.379$\pm$0.340 & 1.920$\pm$0.043 &    $520.7_{-11.3}^{+11.8}$ &      $525.5_{-9.7}^{+9.5}$ \\
            L1322 & 126.566 &  -0.927 &  -530 &   715 &  -14 &      19 &  -1.587$\pm$0.405 &  -1.628$\pm$0.449 & 1.177$\pm$0.037 &    $849.9_{-26.1}^{+27.8}$ &    $890.5_{-21.2}^{+20.8}$ \\
            L1340 & 130.179 &  11.386 &  -551 &   653 &  172 &      25 &  -1.474$\pm$0.202 &  -1.609$\pm$0.228 & 1.091$\pm$0.050 &    $916.3_{-40.3}^{+44.1}$ &    $871.2_{-20.8}^{+22.0}$ \\
            L1355 & 133.014 &   8.861 &  -594 &   637 &  136 &      27 &  -0.440$\pm$0.531 &  -0.564$\pm$0.796 & 1.240$\pm$0.037 &    $806.4_{-23.2}^{+24.7}$ &    $881.8_{-21.0}^{+19.6}$ \\
     L1617and1622 & 204.571 & -11.858 &  -302 &  -138 &  -70 &      10 &   4.884$\pm$4.288 &  -0.043$\pm$3.139 & 2.964$\pm$0.056 &      $337.4_{-6.3}^{+6.5}$ &      $339.4_{-5.8}^{+7.2}$ \\
             L291 &  11.628 &  -1.680 &  1210 &   249 &  -36 &       9 &   1.259$\pm$0.229 &  -1.171$\pm$0.561 & 0.863$\pm$0.059 &   $1159.1_{-74.1}^{+84.9}$ &   $1235.9_{-72.1}^{+74.6}$ \\
       L978and988 &  90.563 &   2.360 &    -6 &   604 &   25 &       7 &   1.548$\pm$0.467 &  -4.004$\pm$0.486 & 1.674$\pm$0.031 &    $597.5_{-10.9}^{+11.3}$ &      $604.7_{-8.2}^{+8.2}$ \\
           Lagoon &   6.093 &  -1.219 &  1234 &   132 &  -26 &       6 &   1.342$\pm$0.415 &  -1.870$\pm$0.426 & 0.828$\pm$0.034 &   $1207.4_{-47.8}^{+51.9}$ &   $1241.8_{-54.6}^{+54.4}$ \\
           LupusI & 338.922 &  15.797 &   137 &   -53 &   42 &       6 & -14.538$\pm$2.039 & -22.663$\pm$0.495 & 6.572$\pm$0.035 &      $152.2_{-0.8}^{+0.8}$ &      $152.8_{-0.7}^{+0.6}$ \\
         LupusIII & 339.600 &   9.342 &   144 &   -54 &   25 &       9 &  -9.508$\pm$0.547 & -24.035$\pm$0.515 & 6.412$\pm$0.010 &      $156.0_{-0.2}^{+0.2}$ &      $156.1_{-0.3}^{+0.3}$ \\
              M20 &   6.619 &  -0.321 &  1227 &   142 &   -7 &       6 &   0.511$\pm$0.752 &  -1.389$\pm$0.217 & 0.827$\pm$0.067 &  $1209.9_{-90.8}^{+106.9}$ &   $1235.7_{-72.9}^{+88.0}$ \\
        Maddalena & 217.071 &  -0.796 & -1986 & -1500 &  -35 &      41 &  -1.181$\pm$0.585 &   0.484$\pm$0.478 & 0.546$\pm$0.076 & $1830.7_{-222.7}^{+294.3}$ & $2489.3_{-181.6}^{+193.5}$ \\
           MonOB1 & 202.853 &   1.856 &  -643 &  -271 &   23 &     142 &  -1.966$\pm$0.504 &  -3.709$\pm$0.490 & 1.456$\pm$0.087 &    $686.8_{-38.7}^{+43.6}$ &      $698.6_{-7.2}^{+7.1}$ \\
            MonR2 & 213.830 & -12.098 &  -679 &  -455 & -175 &      54 &  -3.141$\pm$0.690 &   0.618$\pm$0.460 & 1.265$\pm$0.026 &    $790.5_{-16.2}^{+16.9}$ &    $836.1_{-15.9}^{+15.9}$ \\
          NGC2362 & 238.792 &  -4.338 &  -652 & -1076 &  -95 &      27 &  -2.234$\pm$0.369 &   2.480$\pm$0.445 & 0.747$\pm$0.058 &  $1338.5_{-96.8}^{+113.2}$ &   $1261.5_{-49.7}^{+55.5}$ \\
          NGC6334 & 351.211 &   0.668 &  1701 &  -263 &   20 &      11 &  -0.042$\pm$0.447 &  -1.748$\pm$0.495 & 0.619$\pm$0.067 & $1614.3_{-156.6}^{+194.2}$ & $1721.6_{-111.4}^{+120.3}$ \\
          NGC6604 &  18.324 &   1.979 &  1795 &   595 &   65 &      13 &  -0.456$\pm$0.469 &  -2.156$\pm$0.583 & 0.570$\pm$0.022 &   $1754.0_{-64.1}^{+69.2}$ &   $1892.4_{-78.2}^{+82.8}$ \\
   North\_America &  84.719 &  -0.184 &    78 &   841 &   -3 &      94 &  -1.504$\pm$0.891 &  -3.434$\pm$0.792 & 1.290$\pm$0.031 &    $775.1_{-18.1}^{+19.0}$ &    $845.0_{-25.2}^{+25.9}$ \\
        Oph-L1688 & 353.148 &  16.785 &   132 &   -16 &   40 &      19 &  -7.042$\pm$1.121 & -26.022$\pm$1.421 & 7.277$\pm$0.070 &      $137.4_{-1.3}^{+1.3}$ &      $138.4_{-1.3}^{+1.3}$ \\
         Orion\_A & 210.151 & -19.517 &  -316 &  -184 & -130 &     486 &   0.956$\pm$0.656 &  -0.094$\pm$0.850 & 2.588$\pm$0.024 &      $386.4_{-3.6}^{+3.7}$ &      $388.2_{-2.0}^{+2.0}$ \\
         Orion\_B & 206.018 & -15.372 &  -357 &  -174 & -109 &     106 &   0.084$\pm$1.776 &  -0.991$\pm$1.770 & 2.594$\pm$0.039 &      $385.5_{-5.7}^{+5.9}$ &      $411.6_{-8.2}^{+8.2}$ \\
       Orion\_Lam & 193.905 & -12.497 &  -370 &   -92 &  -84 &      35 &   1.959$\pm$0.816 &  -2.321$\pm$1.299 & 2.538$\pm$0.037 &      $394.0_{-5.7}^{+5.9}$ &      $390.2_{-2.2}^{+2.1}$ \\
     Perseus-East & 160.393 & -18.035 &  -286 &   102 &  -99 &      56 &   4.662$\pm$2.075 &  -6.376$\pm$1.550 & 3.273$\pm$0.054 &      $305.5_{-4.9}^{+5.1}$ &      $319.3_{-9.3}^{+9.0}$ \\
     Perseus-West & 158.303 & -20.616 &  -251 &   100 & -102 &      19 &   7.528$\pm$0.763 &  -9.738$\pm$0.552 & 3.486$\pm$0.017 &      $286.9_{-1.4}^{+1.4}$ &      $288.6_{-1.5}^{+1.6}$ \\
             Pipe & 357.131 &   6.998 &   165 &    -8 &   20 &       6 &  -0.606$\pm$0.798 & -19.162$\pm$0.704 & 6.201$\pm$0.148 &      $161.3_{-3.8}^{+4.0}$ &      $166.4_{-3.9}^{+3.3}$ \\
             RCrA & 359.853 & -17.729 &   144 &     0 &  -46 &       6 &   4.271$\pm$0.763 & -27.447$\pm$0.610 & 6.664$\pm$0.150 &      $150.1_{-3.3}^{+3.4}$ &      $151.1_{-2.3}^{+2.6}$ \\
          Rosette & 207.059 &  -1.980 & -1211 &  -619 &  -47 &      94 &  -1.368$\pm$1.472 &  -0.217$\pm$1.625 & 0.867$\pm$0.052 &   $1152.8_{-65.7}^{+74.1}$ &   $1361.1_{-50.1}^{+55.8}$ \\
             S140 & 107.201 &   5.070 &  -270 &   871 &   81 &      34 &  -1.067$\pm$1.376 &  -2.925$\pm$1.948 & 1.433$\pm$0.166 &    $697.9_{-72.5}^{+91.5}$ &    $915.3_{-61.9}^{+61.9}$ \\
             S235 & 173.467 &   2.626 & -1387 &   159 &   64 &      24 &   0.482$\pm$1.221 &  -2.309$\pm$1.631 & 0.917$\pm$0.109 & $1090.8_{-116.1}^{+147.5}$ &   $1397.2_{-83.7}^{+78.0}$ \\
          Serpens &  31.126 &   4.791 &   380 &   230 &   37 &      25 &   2.845$\pm$0.869 &  -8.876$\pm$0.698 & 2.184$\pm$0.065 &    $457.8_{-13.3}^{+14.1}$ &      $445.6_{-7.3}^{+7.7}$ \\
Taurus-B213-L1495 & 168.806 & -16.018 &  -124 &    24 &  -36 &       9 &   8.735$\pm$0.619 & -24.990$\pm$1.082 & 7.491$\pm$0.084 &      $133.5_{-1.5}^{+1.5}$ &      $131.1_{-1.1}^{+1.2}$ \\
      Taurus-HCL2 & 173.993 & -13.745 &  -134 &    14 &  -33 &       9 &   5.760$\pm$0.667 & -21.117$\pm$0.958 & 7.301$\pm$0.081 &      $137.0_{-1.5}^{+1.5}$ &      $138.4_{-1.2}^{+1.1}$ \\
            VelaA & 270.322 &   0.689 &     9 & -1675 &   20 &      29 &  -5.165$\pm$0.575 &   4.194$\pm$0.531 & 0.506$\pm$0.067 & $1977.7_{-232.2}^{+303.5}$ & $1674.9_{-113.4}^{+123.5}$ \\
            VelaB & 269.614 &  -1.442 &   -12 & -1794 &  -45 &      71 &  -4.790$\pm$3.245 &   4.237$\pm$2.764 & 0.390$\pm$0.099 & $2565.0_{-520.0}^{+874.7}$ & $1794.4_{-121.5}^{+137.6}$ \\
            VelaC & 265.199 &   1.296 &   -83 &  -986 &   22 &      68 &  -6.197$\pm$1.112 &   4.239$\pm$0.660 & 1.114$\pm$0.027 &    $897.6_{-21.5}^{+22.6}$ &    $989.6_{-31.9}^{+30.5}$ \\
            VelaD & 261.947 &   0.539 &  -162 & -1148 &   11 &      97 &  -5.289$\pm$0.969 &   4.091$\pm$0.918 & 1.032$\pm$0.043 &    $969.0_{-39.0}^{+42.4}$ &   $1159.0_{-46.8}^{+49.0}$ \\
               W3 & 133.734 &   0.742 & -1266 &  1323 &   24 &      64 &  -0.735$\pm$1.010 &  -0.299$\pm$0.774 & 0.576$\pm$0.090 & $1735.7_{-234.6}^{+321.5}$ & $1831.2_{-103.1}^{+115.5}$ \\
               W4 & 134.906 &   0.838 & -1259 &  1263 &   26 &      18 &  -0.736$\pm$1.075 &  -0.370$\pm$0.872 & 0.646$\pm$0.053 & $1548.5_{-118.1}^{+139.4}$ & $1784.0_{-111.3}^{+129.6}$ \\
               W5 & 137.065 &   1.202 & -1187 &  1104 &   34 &      63 &  -0.337$\pm$0.980 &  -0.316$\pm$1.186 & 0.748$\pm$0.095 & $1337.2_{-151.2}^{+195.3}$ &  $1621.4_{-91.6}^{+104.8}$ \\
\enddata
\tablecomments{\mz{Columns are: (1) cloud name; (2)-(3) average Galactic coordinates of YSOCs in the clouds; (4)-(6) heliocentric positions of the clouds; (7) number of YSOCs in each cloud; (8)-(9) average proper motions of YSOCs in each cloud. Uncertainties are the standard deviations of proper motions of YSOCs in the clouds; (10) median parallaxes and uncertainties of the local clouds obtained by modelling the parallaxes of YSOCs in clouds with Weibull distribution; (11) median distance obtained as inversion of median parallax; (12) median distances and uncertainties of the local clouds obtained with {\it Kalkayotl}. The machine-readable table can be derived online in the
ChinaVO PaperData repository: \mzac{doi:} \href{https://doi.org/10.12149/101211}{\mzac{10.12149/101210}}.}}
\end{deluxetable*}
\end{longrotatetable}

\section{Results and discussion}\label{sec:discussion}
In Section~\ref{sec:clouddistcat}, I present a catalog of distances to \mz{63} local clouds, constructed using the parallaxes of YSOCs from {\it Gaia} \mz{DR3}. Additionally, I release an all-sky YSOC sample, consisting of \mz{24,883} YSOCs, in Section~\ref{sec:cleanyso}. In this section, I compare our cloud distances with distance estimates from previous literature in Section~\ref{sec:comp2odist}, and also discuss the advantages and limitations of my distance estimates in Section~\ref{sec:caveats}.

\subsection{Comparisons with previous cloud distance estimates}\label{sec:comp2odist}

\begin{figure*}
    \centering
    \includegraphics[width=0.9\linewidth]{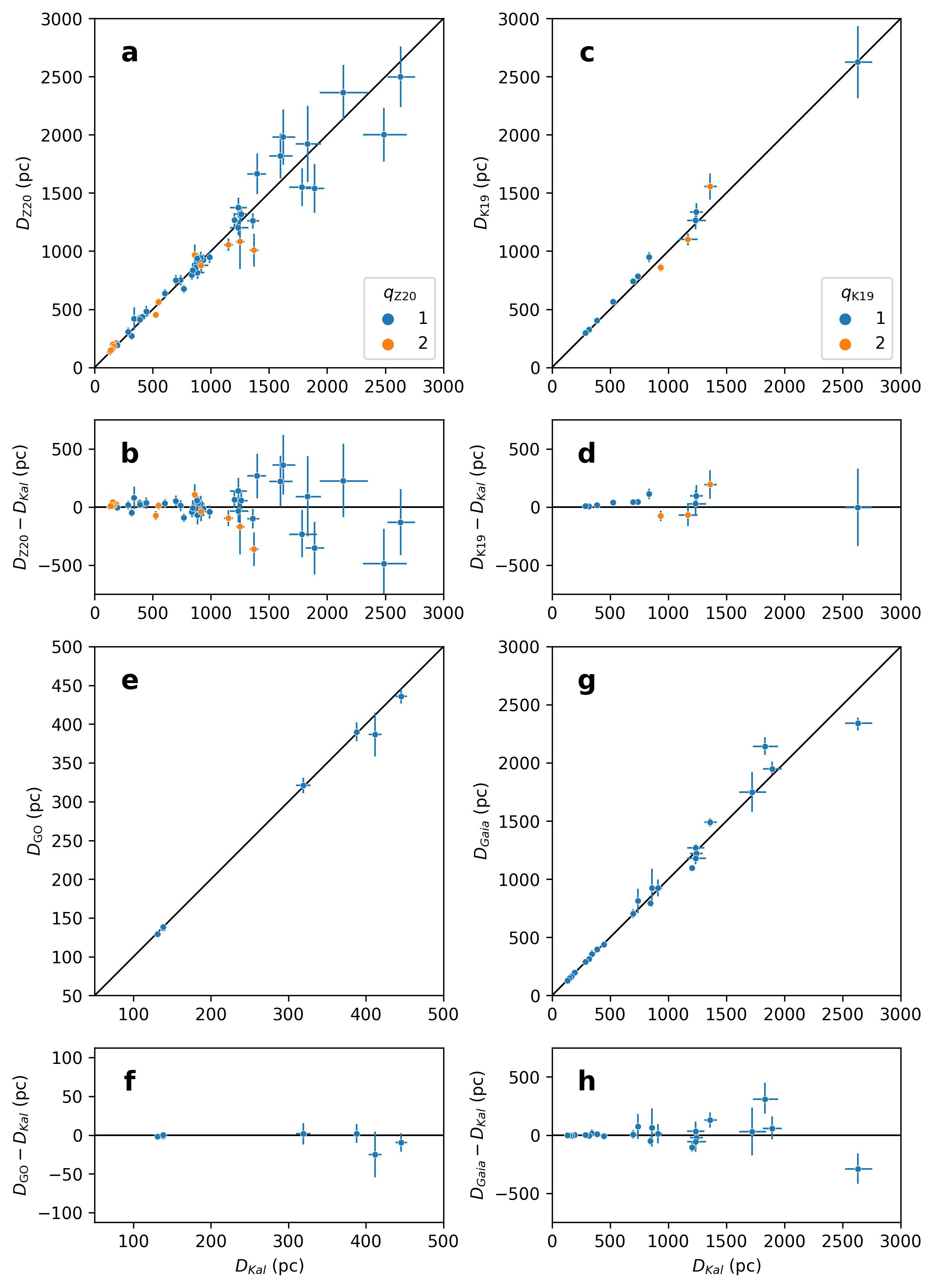}
    \caption{\mz{Comparisons of my distances ($D_{\mathrm{Kal}}$) derived from YSOCs with the average dust Bayesian distances ($D_{\mathrm{Z20}}$) from \citet{zucker2020} (\mztwo{a and b} panels), the average distances ($D_{\mathrm{K19}}$) of young clusters from \citet{kuhn2019} (\mztwo{c and d} panels), the mean distances presented by the GOBELINS project based on VLBI observations (\mztwo{e and f} panels), \mztwo{and the average distances presented by the previous studies based on {\it Gaia} astrometry (g and h panels)} towards the nearby molecular clouds, respectively. The black lines in panels a, c,\mztwo{ e, and g} mark the one-to-one relation while that in panels b, d, \mztwo{f, and h} label the zero difference.}}
    \label{fig:distancecomparisons}
\end{figure*}

First, I compare my catalog of cloud distances ($D_{\mathrm{Kal}}$) presented in Table~\ref{tab:clouddist} with the distance estimates provided by \citet{zucker2020}. To ensure a fair comparison, I calculate a median distance ($D_{\mathrm{Z20}}$) for each local cloud by averaging the distances of sightlines located inside: 1) the cloud boundary defined with $A_V=$~2 mag in Sect.~\ref{sect:sample} ($q_{\mathrm{Z20}}=$~1); or 2) the \mz{whole cloud area} if there is no sightlines inside the cloud boundary ($q_{\mathrm{Z20}}=$~2). The resulting values of $D_{\mathrm{Z20}}$ and $q_{\mathrm{Z20}}$ for 51 local clouds are listed in Table~\ref{tab:clouddistref}. As an example, the dashed orange circles in Fig.~\ref{fig:oriona_dist}a indicate the locations of sightline beams from \citet{zucker2020} in the Orion A cloud. The relations of $D_{\mathrm{Z20}}$ and $D_{\mathrm{Z20}}-D_{\mathrm{Kal}}$ versus $D_{\mathrm{Kal}}$ are presented in Fig.~\ref{fig:distancecomparisons}a and b, respectively. Overall, there is \mz{relatively} good agreement between $D_{\mathrm{Z20}}$ and $D_{\mathrm{Kal}}$. The median and standard deviation of $\frac{D_{\mathrm{Z20}}-D_{\mathrm{Kal}}}{D_{\mathrm{Z20}}}$ are 2\% and \mz{11\%}, respectively. By comparison, the typical errors of $D_{\mathrm{Kal}}$ and $D_{\mathrm{Z20}}$ are \mz{$\sim$3\% and $\sim$7\%}, respectively. Therefore, I find no systematic difference between \mz{$D_{\mathrm{Kal}}$} and $D_{\mathrm{Z20}}$ and \mz{$D_{\mathrm{Kal}}$} is consistent with $D_{\mathrm{Z20}}$ within a typical scatter of \mz{$\sim$11\%}.

Second, I compare $D_{\mathrm{Kal}}$ with the distances of young clusters presented by \citet{kuhn2019}. Using {\it Gaia} DR2 data, \citet{kuhn2019} obtained the median distances of 28 young clusters with ages of $\sim$1$-$5 Myr by averaging the parallax distances of their members. As mentioned above, I also calculate a median distance ($D_{\mathrm{K19}}$) for each local cloud by averaging the distances of young clusters located inside the cloud boundary ($q_{\mathrm{K19}}=$~1) or the entire cloud area ($q_{\mathrm{K19}}=$~2). Table~\ref{tab:clouddistref} lists the $D_{\mathrm{K19}}$ values for 13 local clouds, and the solid \mz{orange} circles in Fig.~\ref{fig:oriona_dist}a indicate the young clusters used to determine the median distance of Orion A. Figure~\ref{fig:distancecomparisons}c and d illustrate the relationship between $D_{\mathrm{K19}}$ and \mz{$D_{\mathrm{Kal}}$} for 13 nearby clouds, indicating good agreement between them. The median and standard deviation of \mz{$\frac{D_{\mathrm{K19}}-D_{\mathrm{Kal}}}{D_{\mathrm{K19}}}$ are approximately 3\% and 5\%}, respectively. Considering the typical error of $D_{\mathrm{K19}}$ (4\%), \mz{$D_{\mathrm{Kal}}$} is consistent with $D_{\mathrm{K19}}$.

Third, I compare $D_{\mathrm{Kal}}$ with the cloud distances provided by the GOBELINS project \citep{gobelins2011,loinard2013}. Table~\ref{tab:clouddistref} shows the GOBELINS distances ($D_{\mathrm{GO}}$) to seven nearby clouds, and Figure~\ref{fig:distancecomparisons}e and f plot the relation between $D_{\mathrm{GO}}$ and $D_{\mathrm{Kal}}$ for these clouds. Given the typical uncertainty ($\sim$2\%) of $D_{\mathrm{GO}}$, the comparison reveals good agreement between the two distance estimates, with a median difference of $<$1\% and a standard deviation of $<$3\% in $\frac{D_{\mathrm{GO}}-D_{\mathrm{Kal}}}{D_{\mathrm{GO}}}$. 

\mztwo{Finally, I compare $D_{\mathrm{Kal}}$ with the {\it Gaia} distances ($D_{Gaia}$) of 26 nearby clouds that were obtained in previous studies. Table~\ref{tab:clouddistref} displays $D_{Gaia}$ and their corresponding references. These distance estimates were obtained using different {\it Gaia} data releases and techniques. Some clouds, such as Taurus and Lupus, have several distance estimates from different studies \citep{luhman2018,luhman2023,galli2019,galli2020,luhman2020}. For these clouds, I generally use the most recent distance estimates with their uncertainties. More studies can be found in \citet{zuckerpp7} (and the references therein). Figure~\ref{fig:distancecomparisons}g and h illustrate the relation between $D_{Gaia}$ and $D_{\mathrm{Kal}}$ for these 26 nearby clouds, indicating good agreement between them. The median and standard deviation of $\frac{D_{Gaia}-D_{\mathrm{Kal}}}{D_{Gaia}}$ are $\sim$1\% and $\sim$5\%, respectively. Considering the typical uncertainty ($\sim$2\%) of $D_{Gaia}$, $D_{\mathrm{Kal}}$ is in close agreement with $D_{Gaia}$.}

I also note that \mz{\mztwo{eight} of the local clouds do not have distance estimates available from \citet{zucker2020}, \citet{kuhn2019}, the GOBELINS project, or previous case studies based on {\it Gaia} data.} Therefore, I have gathered the \mztwo{other available} distance estimates ($D_{\mathrm{lit}}$) for them from literature. Table~\ref{tab:clouddistref} provides a list of $D_{\mathrm{lit}}$ and corresponding references. For most of these \mztwo{eight} clouds, $D_{\mathrm{Kal}}$ agrees reasonably well with $D_{\mathrm{lit}}$, with the exception of Vela A. The $D_{\mathrm{Kal}}$ value for Vela A is $1674.9_{-113.4}^{+123.5}$ 
which is roughly twice the value of $D_{\mathrm{lit}}=$~700$\pm$200 pc reported by \citet{liseau1992}. 

\citet{liseau1992} estimated the distance of the Vela Molecular Ridge (VMR) that can be divided into four regions, i.e., Vela A-D \citep{murphy1991}, based on 1) the photometric distances of some infrared sources; 2) star counts distances; and 3) the distances of reflection nebulae and OB associations from literature. Finally they suggested that Vela ACD have a similar distance of 0.7$\pm$0.2 kpc while Vela B has a larger distance of $\sim$2 kpc. Recently, \citet{hottier2021} obtained the 3D extinction density of the Vela complex using the Field Extinction-Distance Relation Deconvolver \citep[FEDReD,][]{fedred1,fedred2} algorithm based on the 2MASS and {\it Gaia} DR2 data. They found that the Vela C extended from 0.47 kpc to 1.66 kpc and its barycenter was at the distance of 0.90$\pm$0.09 kpc, which was consistent with my result of \mztwo{$D_{\mathrm{Kal}}=989.6_{-31.9}^{+30.5}$ pc} and that of $D_{\mathrm{Z20}}=$0.947$\pm$0.048 kpc from \citet{zucker2020}. \citet{hottier2021} also found that the Vela D extended from 0.75 kpc to 3.30 kpc with a barycenter distance of 1.90$\pm$0.38 kpc, which inferred that \citet{liseau1992} and my work only obtained the distance of the front part of Vela D. Especially, \citet{hottier2021} suggested that the Vela A and B were actually one unique structure that extended from 1.13 kpc to 3.37 kpc with a barycenter distance of 2.17$\pm$0.01 kpc, which contradicted \citet{liseau1992}'s results. However, \mztwo{my results of $D_{\mathrm{Kal,Vela A}}=1674.9_{-113.4}^{+123.5}$ pc and $D_{\mathrm{Kal,Vela B}}=1794.4_{-121.5}^{+137.6}$ pc} are consistent with \citet{hottier2021}'s result if assuming that we only see the front part of Vela A and B.

In general, my results are in good agreement with previous distance estimates from the literature. This provides strong support for the accuracy of my cloud distances, particularly for those clouds that lack {\it Gaia}-based distance measurements, such as the Cartwheel and Vela AB clouds.

\begin{longrotatetable}
\begin{deluxetable*}{lcccccrcrcr}
\tabletypesize{\scriptsize}
\tablecaption{Distances from literature for \mz{63} nearby molecular clouds\label{tab:clouddistref}}
\tablehead{
\colhead{Cloud} & \colhead{$D_{\mathrm{Z20}}$} & \colhead{$q_{\mathrm{Z20}}$} & \colhead{$D_{\mathrm{K19}}$} & \colhead{$q_{\mathrm{K19}}$} & 
\colhead{$D_{\mathrm{GO}}$} &
\colhead{Ref$_{\mathrm{GO}}$}&
\colhead{$D_{\mathrm{Gaia}}$}&
\colhead{Ref$_{\mathrm{Gaia}}$}&
\colhead{$D_{\mathrm{lit}}$} & \colhead{Ref$_{\mathrm{lit}}$} \\
\colhead{} & \colhead{(pc)} & \colhead{} & \colhead{(pc)} & \colhead{} & \colhead{(pc)}&\colhead{} & \colhead{(pc)} & \colhead{} & \colhead{pc} &
\colhead{}
}
\colnumbers
\startdata
AFGL490 &     $\ldots$ &   $\ldots$ &     $\ldots$ &   $\ldots$ &          $\ldots$ &                 $\ldots$ &                   $\ldots$ &              $\ldots$ & 1000$\pm$500 &   \citet{snell1984} \\
              Ara &  1055$\pm$55 &          2 &     $\ldots$ &   $\ldots$ &          $\ldots$ &                 $\ldots$ &                   $\ldots$ &              $\ldots$ &     $\ldots$ &            $\ldots$ \\
         CMa\_OB1 &  1266$\pm$63 &          1 &     $\ldots$ &   $\ldots$ &          $\ldots$ &                 $\ldots$ &   $1099.0_{-24.0}^{+25.0}$ &        \citet{gh2021} &     $\ldots$ &            $\ldots$ \\
       California &   452$\pm$29 &          2 &   564$\pm$15 &          1 &          $\ldots$ &                 $\ldots$ &                   $\ldots$ &              $\ldots$ &     $\ldots$ &            $\ldots$ \\
           Carina & 2498$\pm$260 &          1 & 2626$\pm$310 &          1 &          $\ldots$ &                 $\ldots$ &   $2340.0_{-60.0}^{+50.0}$ &     \citet{goppl2022} &     $\ldots$ &            $\ldots$ \\
        Cartwheel &     $\ldots$ &   $\ldots$ &     $\ldots$ &   $\ldots$ &          $\ldots$ &                 $\ldots$ &                   $\ldots$ &              $\ldots$ &         1800 &   \citet{lopez2011} \\
         Cep\_OB3 &     $\ldots$ &   $\ldots$ &   859$\pm$33 &          2 &          $\ldots$ &                 $\ldots$ &                   $\ldots$ &              $\ldots$ &     $\ldots$ &            $\ldots$ \\
         Cep\_OB4 &     $\ldots$ &   $\ldots$ &  1100$\pm$50 &          2 &          $\ldots$ &                 $\ldots$ &                   $\ldots$ &              $\ldots$ &     $\ldots$ &            $\ldots$ \\
    Cepheus-L1251 &   421$\pm$96 &          1 &     $\ldots$ &   $\ldots$ &          $\ldots$ &                 $\ldots$ &    $358.0_{-32.0}^{+32.0}$ &      \citet{dzib2018} &     $\ldots$ &            $\ldots$ \\
  Cepheus-NGC7129 &   966$\pm$89 &          2 &     $\ldots$ &   $\ldots$ &          $\ldots$ &                 $\ldots$ &  $926.0_{-163.0}^{+163.0}$ &      \citet{dzib2018} &     $\ldots$ &            $\ldots$ \\
            ChamI &   210$\pm$20 &          2 &     $\ldots$ &   $\ldots$ &          $\ldots$ &                 $\ldots$ &      $189.4_{-0.7}^{+0.8}$ &     \citet{galli2021} &     $\ldots$ &            $\ldots$ \\
           ChamII &   190$\pm$14 &          1 &     $\ldots$ &   $\ldots$ &          $\ldots$ &                 $\ldots$ &      $197.5_{-0.9}^{+1.0}$ &     \citet{galli2021} &     $\ldots$ &            $\ldots$ \\
         Circinus &   675$\pm$34 &          1 &     $\ldots$ &   $\ldots$ &          $\ldots$ &                 $\ldots$ &                   $\ldots$ &              $\ldots$ &     $\ldots$ &            $\ldots$ \\
       Crossbones &   923$\pm$48 &          1 &     $\ldots$ &   $\ldots$ &          $\ldots$ &                 $\ldots$ &                   $\ldots$ &              $\ldots$ &     $\ldots$ &            $\ldots$ \\
      Cygnus-West &     $\ldots$ &   $\ldots$ &     $\ldots$ &   $\ldots$ &          $\ldots$ &                 $\ldots$ &                   $\ldots$ &              $\ldots$ &         1800 &   \citet{green2019} \\
    CygnusX-North & 1082$\pm$236 &          2 &     $\ldots$ &   $\ldots$ &          $\ldots$ &                 $\ldots$ &                   $\ldots$ &              $\ldots$ &     $\ldots$ &            $\ldots$ \\
    CygnusX-South & 1008$\pm$141 &          2 &     $\ldots$ &   $\ldots$ &          $\ldots$ &                 $\ldots$ &                   $\ldots$ &              $\ldots$ &     $\ldots$ &            $\ldots$ \\
     G82.65-02.00 &   814$\pm$52 &          1 &     $\ldots$ &   $\ldots$ &          $\ldots$ &                 $\ldots$ &                   $\ldots$ &              $\ldots$ &     $\ldots$ &            $\ldots$ \\
         Gem\_OB1 & 1818$\pm$194 &          1 &     $\ldots$ &   $\ldots$ &          $\ldots$ &                 $\ldots$ &                   $\ldots$ &              $\ldots$ &     $\ldots$ &            $\ldots$ \\
           IC1396 &   941$\pm$55 &          1 &     $\ldots$ &   $\ldots$ &          $\ldots$ &                 $\ldots$ &    $925.0_{-73.0}^{+73.0}$ &        \citet{pb2023} &     $\ldots$ &            $\ldots$ \\
           IC2944 & 2363$\pm$238 &          1 &     $\ldots$ &   $\ldots$ &          $\ldots$ &                 $\ldots$ &                   $\ldots$ &              $\ldots$ &     $\ldots$ &            $\ldots$ \\
           IC5146 &   752$\pm$45 &          1 &   783$\pm$26 &          1 &          $\ldots$ &                 $\ldots$ &  $813.0_{-106.0}^{+106.0}$ &      \citet{dzib2018} &     $\ldots$ &            $\ldots$ \\
     L1003and1004 &   562$\pm$32 &          2 &     $\ldots$ &   $\ldots$ &          $\ldots$ &                 $\ldots$ &                   $\ldots$ &              $\ldots$ &     $\ldots$ &            $\ldots$ \\
            L1265 &     $\ldots$ &   $\ldots$ &     $\ldots$ &   $\ldots$ &          $\ldots$ &                 $\ldots$ &                   $\ldots$ &              $\ldots$ &          600 &     \citet{kun2008} \\
            L1302 &     $\ldots$ &   $\ldots$ &     $\ldots$ &   $\ldots$ &          $\ldots$ &                 $\ldots$ &                   $\ldots$ &              $\ldots$ &          600 &     \citet{kun2008} \\
            L1322 &   922$\pm$53 &          1 &     $\ldots$ &   $\ldots$ &          $\ldots$ &                 $\ldots$ &                   $\ldots$ &              $\ldots$ &     $\ldots$ &            $\ldots$ \\
            L1340 &   858$\pm$44 &          1 &     $\ldots$ &   $\ldots$ &          $\ldots$ &                 $\ldots$ &                   $\ldots$ &              $\ldots$ &     $\ldots$ &            $\ldots$ \\
            L1355 &   936$\pm$52 &          1 &     $\ldots$ &   $\ldots$ &          $\ldots$ &                 $\ldots$ &                   $\ldots$ &              $\ldots$ &     $\ldots$ &            $\ldots$ \\
     L1617and1622 &   418$\pm$27 &          1 &     $\ldots$ &   $\ldots$ &          $\ldots$ &                 $\ldots$ &                   $\ldots$ &              $\ldots$ &     $\ldots$ &            $\ldots$ \\
             L291 &  1374$\pm$85 &          1 &     $\ldots$ &   $\ldots$ &          $\ldots$ &                 $\ldots$ &   $1270.0_{-30.0}^{+30.0}$ &      \citet{kuhn2021} &     $\ldots$ &            $\ldots$ \\
       L978and988 &   635$\pm$38 &          1 &     $\ldots$ &   $\ldots$ &          $\ldots$ &                 $\ldots$ &                   $\ldots$ &              $\ldots$ &     $\ldots$ &            $\ldots$ \\
           Lagoon &     $\ldots$ &   $\ldots$ &  1336$\pm$76 &          1 &          $\ldots$ &                 $\ldots$ &   $1220.0_{-20.0}^{+10.0}$ &      \citet{kuhn2021} &     $\ldots$ &            $\ldots$ \\
           LupusI &   156$\pm$14 &          1 &     $\ldots$ &   $\ldots$ &          $\ldots$ &                 $\ldots$ &      $154.9_{-3.4}^{+3.2}$ &     \citet{galli2020} &     $\ldots$ &            $\ldots$ \\
         LupusIII &   197$\pm$15 &          2 &     $\ldots$ &   $\ldots$ &          $\ldots$ &                 $\ldots$ &      $158.9_{-0.7}^{+0.7}$ &     \citet{galli2020} &     $\ldots$ &            $\ldots$ \\
              M20 &  1201$\pm$66 &          1 &  1264$\pm$76 &          1 &          $\ldots$ &                 $\ldots$ &   $1180.0_{-50.0}^{+50.0}$ &      \citet{kuhn2021} &     $\ldots$ &            $\ldots$ \\
        Maddalena & 2001$\pm$232 &          1 &     $\ldots$ &   $\ldots$ &          $\ldots$ &                 $\ldots$ &                   $\ldots$ &              $\ldots$ &     $\ldots$ &            $\ldots$ \\
           MonOB1 &   749$\pm$47 &          1 &   740$\pm$24 &          1 &          $\ldots$ &                 $\ldots$ &    $704.0_{-38.0}^{+38.0}$ &       \citet{lim2022} &     $\ldots$ &            $\ldots$ \\
            MonR2 &   793$\pm$41 &          1 &   948$\pm$42 &          1 &          $\ldots$ &                 $\ldots$ &                   $\ldots$ &              $\ldots$ &     $\ldots$ &            $\ldots$ \\
          NGC2362 &  1317$\pm$66 &          1 &     $\ldots$ &   $\ldots$ &          $\ldots$ &                 $\ldots$ &                   $\ldots$ &              $\ldots$ &     $\ldots$ &            $\ldots$ \\
          NGC6334 &     $\ldots$ &   $\ldots$ &     $\ldots$ &   $\ldots$ &          $\ldots$ &                 $\ldots$ & $1750.0_{-170.0}^{+170.0}$ &   \citet{russeil2020} & 1740$\pm$310 & \citet{ngc6334dist} \\
          NGC6604 & 1539$\pm$211 &          1 &     $\ldots$ &   $\ldots$ &          $\ldots$ &                 $\ldots$ &   $1950.0_{-50.0}^{+60.0}$ &      \citet{kuhn2021} &     $\ldots$ &            $\ldots$ \\
   North\_America &   835$\pm$63 &          1 &     $\ldots$ &   $\ldots$ &          $\ldots$ &                 $\ldots$ &    $795.0_{-25.0}^{+25.0}$ &      \citet{kuhn2020} &     $\ldots$ &            $\ldots$ \\
        Oph-L1688 &    139$\pm$8 &          1 &     $\ldots$ &   $\ldots$ &     137.3$\pm$1.2 &     \citet{gobelins-oph} &                   $\ldots$ &              $\ldots$ &     $\ldots$ &            $\ldots$ \\
         Orion\_A &   423$\pm$34 &          1 &    403$\pm$7 &          1 &    390.1$\pm$12.0\tablenotemark{a} &   \citet{gobelins-orion} &    $397.0_{-16.0}^{+16.0}$ &  \citet{oriona3d2018} &     $\ldots$ &            $\ldots$ \\
         Orion\_B &   434$\pm$31 &          1 &     $\ldots$ &   $\ldots$ &    386.6$\pm$28.2\tablenotemark{a} &   \citet{gobelins-orion} &                   $\ldots$ &              $\ldots$ &     $\ldots$ &            $\ldots$ \\
       Orion\_Lam &   409$\pm$27 &          1 &     $\ldots$ &   $\ldots$ &          $\ldots$ &                 $\ldots$ &                   $\ldots$ &              $\ldots$ &     $\ldots$ &            $\ldots$ \\
     Perseus-East &   270$\pm$32 &          1 &    324$\pm$5 &          1 &    321.0$\pm$10.0 & \citet{gobelins-perseus} &      $315.0_{-1.0}^{+1.0}$ &  \citet{olivares2022} &     $\ldots$ &            $\ldots$ \\
     Perseus-West &   305$\pm$38 &          1 &    296$\pm$5 &          1 &          $\ldots$ &                 $\ldots$ &      $292.0_{-1.0}^{+1.0}$ &  \citet{olivares2022} &     $\ldots$ &            $\ldots$ \\
             Pipe &   180$\pm$11 &          2 &     $\ldots$ &   $\ldots$ &          $\ldots$ &                 $\ldots$ &      $163.0_{-5.0}^{+5.0}$ &      \citet{dzib2018} &     $\ldots$ &            $\ldots$ \\
             RCrA &   156$\pm$14 &          2 &     $\ldots$ &   $\ldots$ &          $\ldots$ &                 $\ldots$ &      $149.4_{-0.4}^{+0.4}$ & \citet{gallircra2020} &     $\ldots$ &            $\ldots$ \\
          Rosette &  1261$\pm$65 &          1 & 1555$\pm$110 &          2 &          $\ldots$ &                 $\ldots$ &   $1489.0_{-37.0}^{+37.0}$ &     \citet{muzic2022} &     $\ldots$ &            $\ldots$ \\
             S140 &   876$\pm$53 &          2 &     $\ldots$ &   $\ldots$ &          $\ldots$ &                 $\ldots$ &                   $\ldots$ &              $\ldots$ &     $\ldots$ &            $\ldots$ \\
             S235 & 1665$\pm$173 &          1 &     $\ldots$ &   $\ldots$ &          $\ldots$ &                 $\ldots$ &                   $\ldots$ &              $\ldots$ &     $\ldots$ &            $\ldots$ \\
          Serpens &   481$\pm$48 &          1 &     $\ldots$ &   $\ldots$ &     436.0$\pm$9.2 & \citet{gobelins-serpens} &    $438.0_{-11.0}^{+11.0}$ &      \citet{greg2019} &     $\ldots$ &            $\ldots$ \\
Taurus-B213-L1495 &   134$\pm$10 &          2 &     $\ldots$ &   $\ldots$ &     129.5$\pm$0.3 &  \citet{gobelins-taurus} &      $129.9_{-0.3}^{+0.4}$ &     \citet{galli2019} &     $\ldots$ &            $\ldots$ \\
      Taurus-HCL2 &   147$\pm$13 &          2 &     $\ldots$ &   $\ldots$ &     138.6$\pm$2.1 &  \citet{gobelins-taurus} &                   $\ldots$ &              $\ldots$ &     $\ldots$ &            $\ldots$ \\
            VelaA &     $\ldots$ &   $\ldots$ &     $\ldots$ &   $\ldots$ &          $\ldots$ &                 $\ldots$ &                   $\ldots$ &              $\ldots$ &  700$\pm$200 &  \citet{liseau1992} \\
            VelaB &     $\ldots$ &   $\ldots$ &     $\ldots$ &   $\ldots$ &          $\ldots$ &                 $\ldots$ &                   $\ldots$ &              $\ldots$ &         2000 &  \citet{liseau1992} \\
            VelaC &   947$\pm$48 &          1 &     $\ldots$ &   $\ldots$ &          $\ldots$ &                 $\ldots$ &                   $\ldots$ &              $\ldots$ &     $\ldots$ &            $\ldots$ \\
            VelaD &     $\ldots$ &   $\ldots$ &     $\ldots$ &   $\ldots$ &          $\ldots$ &                 $\ldots$ &                   $\ldots$ &              $\ldots$ &  700$\pm$200 &  \citet{liseau1992} \\
               W3 & 1921$\pm$327 &          1 &     $\ldots$ &   $\ldots$ &          $\ldots$ &                 $\ldots$ &   $2140.0_{-70.0}^{+80.0}$ &  \citet{navarete2019} &     $\ldots$ &            $\ldots$ \\
               W4 & 1549$\pm$163 &          1 &     $\ldots$ &   $\ldots$ &          $\ldots$ &                 $\ldots$ &                   $\ldots$ &              $\ldots$ &     $\ldots$ &            $\ldots$ \\
               W5 & 1981$\pm$238 &          1 &     $\ldots$ &   $\ldots$ &          $\ldots$ &                 $\ldots$ &                   $\ldots$ &              $\ldots$ &     $\ldots$ &            $\ldots$ \\
\enddata
\tablenotetext{a}{\mz{These distances are re-calculated by averaging the published GOBELINS distances of YSOs in my defined cloud boundaries.}}
\tablecomments{\mz{Columns are (1) cloud name; (2) distances from \citet{zucker2020}; (3) quality flag (see text for details); (4) distances from \citet{kuhn2019}; (5) quality flag (see text for details); (6) distances from GOBELINS project \citep{gobelins2011} based on the VLBI observations; (7) corresponding references; \mztwo{(8) distances from previous studies based on Gaia astrometry; (9) corresponding references}; (10) distances from other studies; (11) corresponding references of those studies. The machine-readable table can be derived online in the
ChinaVO PaperData repository: \mzac{doi:} \href{https://doi.org/10.12149/101211}{\mzac{10.12149/101210}}.}}

\end{deluxetable*}
\end{longrotatetable}

\subsection{Advantages and caveats of my distance estimates}\label{sec:caveats}

In Section~\ref{sec:comp2odist}, I compared my distance estimates for local clouds with those from the literature, including the distance catalog presented by \citet{zucker2020}. While many studies have investigated the distances to nearby molecular clouds \citep{schlafly2014,dzib2018,yan2019,yan2019b,yan2020,yan2021a,yan2021b,chen2020,guo2022,prisinzano2022}, to the best of my knowledge, \citet{zucker2020}'s catalog is the largest and most homogeneous one to date, providing distances to most of the well-studied local clouds within 2.5 kpc of the Sun. In this section, I discuss the main advantages and limitations of my distance estimates compared to those in \citet{zucker2020}'s catalog.

\mz{The first advantage of my distance catalog is its higher precision. The typical uncertainty of distances in my catalog is about 3\%, which is smaller than the typical error of distances in \citet{zucker2020}'s catalog, which is about 7\%. The improved distance precision can be attributed to the increased parallax accuracy in the {\it Gaia} DR3 catalog, compared to DR2.} 

My distance catalog has a second advantage in that it provides distance estimates for approximately 10 local clouds, including the Cartwheel and Vela ABD, which are not included in \citet{zucker2020}'s catalog. In particular, my results suggest that Vela A and B are located at a similar distance of \mz{$\sim$1.6$-$1.7} kpc, which contradicts the widely-used distance values proposed by \citet{liseau1992}, but is consistent with the structures recently revealed by a 3D extinction density map \citep{fedred2}.

\begin{figure}
    \centering
    \includegraphics[width=1.0\linewidth]{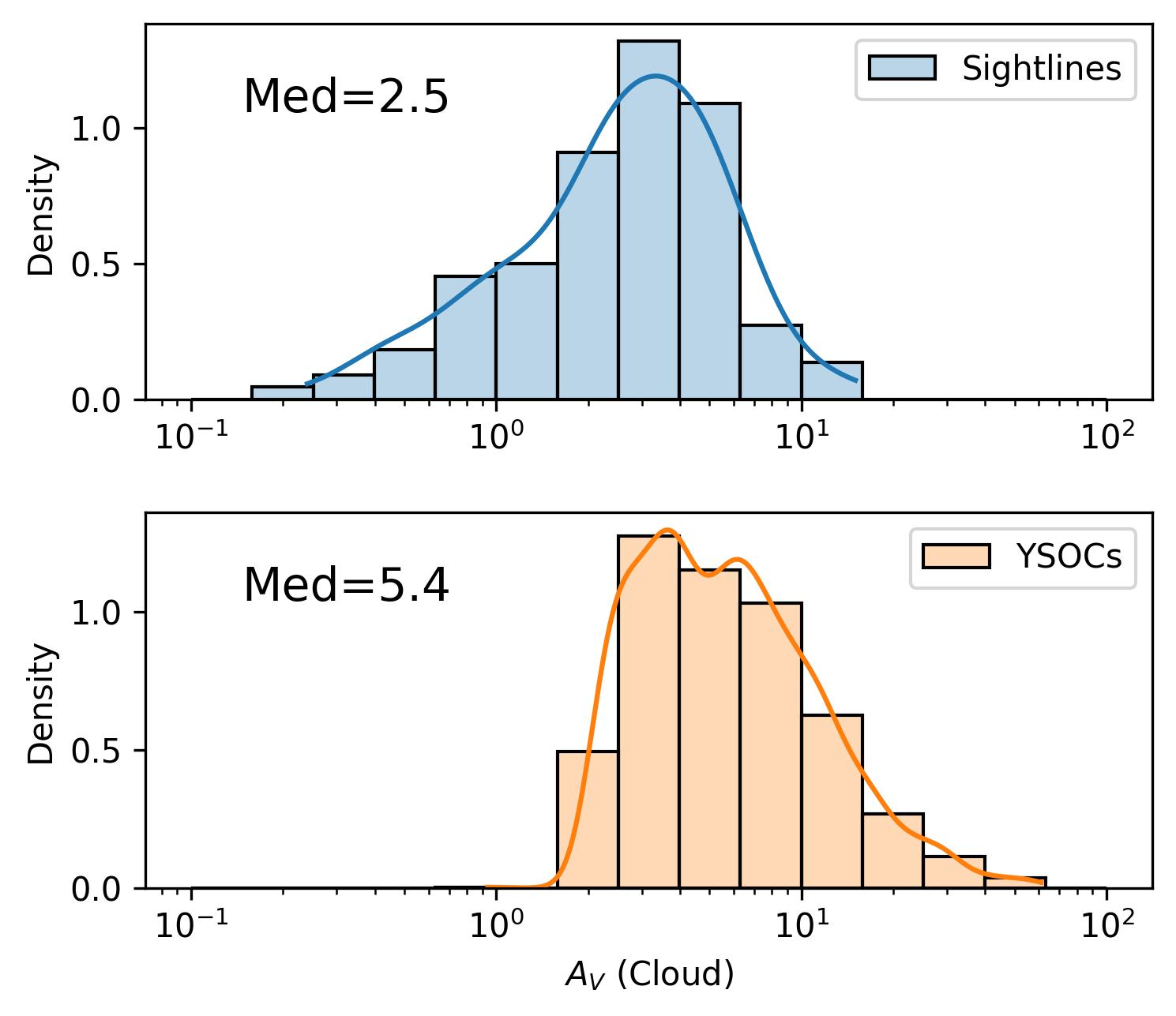}
    \caption{The distributions of cloud column densities, i.e., $A_V$(Cloud), in locations of the sightline beams from \citet{zucker2020}'s catalog (top) and the YSOCs in the local clouds (bottom). The solid lines are the KDEs calculated from the histograms. I also mark the median value of the distribution in each panel.}
    \label{fig:avcloudcomp}
\end{figure}

The third advantage of my distance catalog is that it enables tracing the distances to the denser parts of local clouds, which could be more closely linked to the process of star formation \citep{lada2010,gao2004,mypub2019}. Among the 51 clouds for which both my distance estimates and those of \citet{zucker2020} are available, I extract the cloud column density, $A_V$(Cloud), from 124 sightlines in \citet{zucker2020}'s catalog and from \mztwo{2,\,603} YSOCs selected by me. Figure~\ref{fig:avcloudcomp} illustrates the distributions of $A_V$(Cloud) for sightlines and YSOCs, respectively. A systematic difference between the two distributions can be observed. The median values of $A_V$(Cloud) for sightlines and YSOCs are 2.5 and \mz{5.4} mag, respectively. Therefore, while \citet{zucker2020}'s catalog provides the distances to the relatively low-density regions of nearby clouds, my distances are obtained based on the YSOCs that are located in significantly denser regions of local clouds.

The main disadvantage of my catalog is that it solely provides distances to local clouds with star formation. My methodology used to estimate cloud distances is dependent on the number of YSOs available, therefore, clouds without sufficient YSOs cannot have distances estimated. On the other hand, \citet{zucker2020}'s catalog uses the distances and extinctions of stars, which does not have a similar limitation. Consequently, \citet{zucker2020} is able to present distances to tens of quiescent local clouds, such as Polaris and Coalsack.

\section{Conclusion}\label{sec:summary}
I determine the distances to \mz{63} molecular clouds located within \mz{$\sim$2.5 kpc} of the Sun using the YSOs associated with these clouds and the {\it Gaia} \mz{DR3 parallaxes}. 

I compile an all-sky sample of YSOCs by identifying young stars using the AllWISE catalog, and combining it with previous all-sky YSO catalogs \citep{ysowise2016,ysowise2019}. I exclude a large part of possible contamination in my YSOC sample, such as MS stars, giants, and AGBs, by introducing stellar evolutionary tracks and YSO models \citep{stellarmodel_parsec2012,stellarmodel_colibri2013,ysomodel2006,ysomodel2007}. The resulting clean catalog contains \mz{24,883} YSOCs. By comparing with the previous YSO catalog in Orion A \citep{ysooriona2019}, I estimate that the contamination fraction and completeness of my clean YSOC catalog could be $\sim$30\% and $<$10\%, respectively. 

After removing any outliers with additional astrometric information from {\it Gaia} DR3 using DBSCAN \citep{dbscan1996}, I associate \mztwo{3,\,144} YSOCs with \mz{63} nearby molecular clouds. I then determine precise cloud distances by fitting the distribution of YSOC \mz{parallaxes using the {\it Kalkayotl} code \citep{kalkayotl2020}} for each local cloud. The typical uncertainty in my cloud distance measurements is approximately \mz{3\%}.

My catalog of cloud distances demonstrates good agreement with previous distance estimates \citep{gobelins2011,kuhn2019,zucker2020}, with typical scatters of $\lesssim$10\%. Unlike recent catalogs based on dust extinction \citep{zucker2020,yan2019,chen2020} that provide homogeneous distances to local clouds, my catalog focuses on the relatively dense areas within these clouds. This makes my catalog more appropriate as a reference for investigating the physical properties of nearby dense regions. 

My method primarily depends on YSOs with \mz{parallax} measurements in the molecular clouds. With the upcoming {\it Gaia} data releases and all-sky deep surveys like LSST \citep{lsst2019}, there will be a significant increase in the number of YSOs in the solar neighborhood. This will provide exciting opportunities to map the 3D structures of local clouds, particularly the dense regions.


\begin{acknowledgments}
This work was supported by the National Natural Science Foundation of China (grants No. 12073079). This publication makes use of data products from the Wide-field
Infrared Survey Explorer, which is a joint project of the University of
California, Los Angeles, and the Jet Propulsion Laboratory/California
Institute of Technology, funded by the National Aeronautics and Space
Administration. This work has made use of data from the European Space Agency (ESA) mission
{\it Gaia} (\url{https://www.cosmos.esa.int/gaia}), processed by the {\it Gaia}
Data Processing and Analysis Consortium (DPAC,
\url{https://www.cosmos.esa.int/web/gaia/dpac/consortium}). Funding for the DPAC
has been provided by national institutions, in particular the institutions
participating in the {\it Gaia} Multilateral Agreement. This publication makes use of data products from the Two Micron All Sky Survey, which is a joint project of the University of Massachusetts and the Infrared Processing and Analysis Center/California Institute of Technology, funded by the National Aeronautics and Space Administration and the National Science Foundation. This research has made use of the SIMBAD database, operated at CDS, Strasbourg, France. This research made use of Astropy\footnote{\url{http://www.astropy.org}}, a community-developed core Python package for Astronomy \citep{astropy:2013,astropy:2018}.
\end{acknowledgments}

%

\vspace{5mm}
\facilities{WISE, {\it Gaia}, Planck}


\software{astropy \citep{astropy:2013,astropy:2018}, 
          DBSCAN \citep{dbscan1996},
          reliability \citep{reliability},
          Kalkayotl \citep{kalkayotl2020},
          Stilts \citep{stilts2006},
          Sextractor \citep{sex1996}}



\appendix

\section{Estimating extinctions and de-reddened spectral indices of YSOCs}\label{ap:extalphac}

\begin{figure*}
    \centering
    \includegraphics[width=1.0\linewidth]{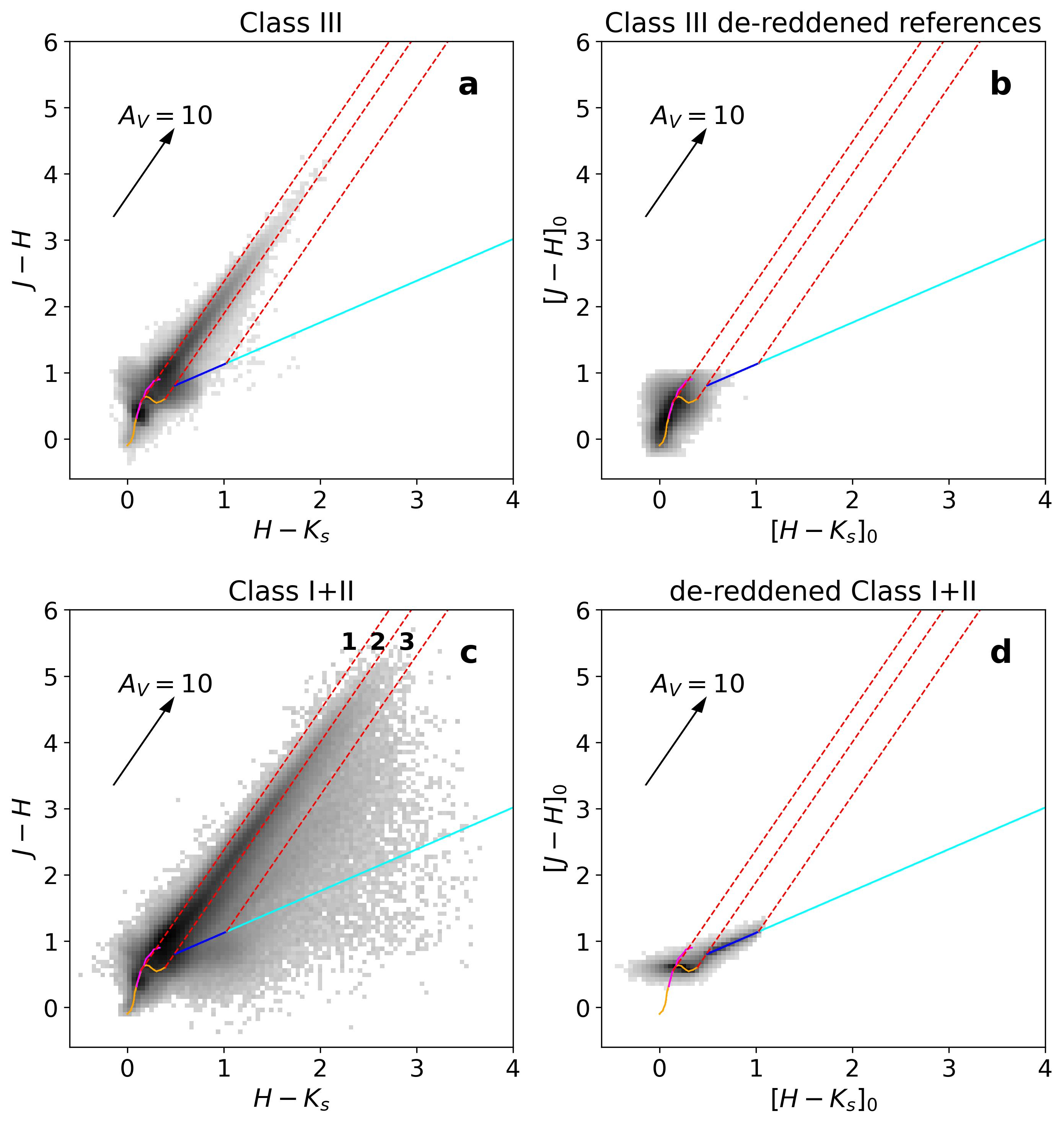}
    \caption{The de-reddening process of the YSOCs. The left two panels show the $H-K_s$ versus $J-H$ CCDs for \mz{Class III} candidates (a) and Class I/II candidates (c), respectively. The right two panels show the de-reddened $[H-K_s]_0$ versus $[J-H]_0$ CCDs for (b): the \mz{Class III} references that are as a input to PNICER method; and (d): the Class I/II candidates that are located in sub-regions 1, 2, and 3 marked in panel c. The solid curves show the intrinsic colors for the main-sequence stars (orange) and giants (magenta), individually \citep{bb1988}. The blue solid lines label the locus of T Tauri stars suggested by \citet{meyer1997} and the cyan lines are the extrapolation of the T Tauri locus. The red dashed lines show the reddening direction and separate the color plane into three sub-regions (1, 2, and 3). I use different methods to estimate the foreground extinctions of YSOCs in different sub-regions (see text for details). The black arrows mark the reddening vectors \citep{wang2019}.}
    \label{fig:avest}
\end{figure*}

I begin by calculating the observed spectral indices ($\alpha$) of YSOCs and classify them into Class I, II, and III candidates using the scheme proposed by \citet{lada1987}. Next, I estimate the foreground extinctions of Class I, II, and III YSOCs using different methods. Then, I re-classify the YSOCs based on their de-reddened spectral indices and re-calculate the extinctions. I repeat this process several times until I obtain non-variable de-reddened spectral indices of the YSOCs. The following are the detailed steps of this iterative process:

\begin{itemize}
    \item[1.]{I calculate $\alpha$ of YSOCs by fitting their observed SEDs from 2 to 22\,\micron, i.e., fluxes in the $K_s$ and $W1-W4$ bands.}
    \item[2.]{The YSOCs with $\alpha<-$2 are isolated as the \mz{Class III} candidates. I find that about \mz{40\%} of the \mz{Class III} candidates have \mz{GSP-Phot} extinction estimates ($A_{G}$). Because the Class \mz{III} sources usually have little infrared excess, their SEDs can be approached with the evolutionary tracks of MS stars. Thus the \mz{$A_G$} values of the Class \mz{III} candidates should be reliable. }
    \item[3.]{For the Class \mz{III} candidates without \mz{$A_{G}$} values, I use PNICER method to estimate their foreground extinctions \mz{($A_{G,\mathrm{PNICER}}$)}. PNICER \citep{pnicer2017} is an unsupervised machine learning technique, which determines the probability distribution of extinction by fitting the features of reference sources in color space. I use the \mz{Class III} candidates with \mz{$A_{G}$} estimates as reference sources. I also apply a cut to the de-reddened color of reference sources, i.e., $[J-H]_0<$~1 mag. Figure~\ref{fig:avest}a shows the $H-K_s$ versus $J-H$ color-color diagram (CCD) for the \mz{Class III} candidates while Fig.~\ref{fig:avest}b shows the de-reddened $[H-K_s]_0$ versus $[J-H]_0$ CCD for the Class \mz{III} candidates with \mz{$A_{G}$} estimates. The extinctions towards the Class \mz{III} candidates without \mz{$A_{G}$} are obtained by comparing their observed colors to the de-reddened colors of the Class \mz{III} candidates with \mz{$A_{G}$} using the PNICER python package\footnote{\url{https://github.com/smeingast/PNICER}} based on the extinction law suggested by \citet{wang2019}. I note that \mz{$\sim$1\%} Class \mz{III} candidates have neither \mz{$A_{G}$} nor positive \mz{$A_{G,\mathrm{PNICER}}$}.}
    \item[4.]{For the Class I/II candidates ($\alpha \geqslant -$2) with $J,H,K_s$ detections, their extinctions are obtained by employing the $JHK_s$ CCD. A detailed description of the scheme can be found in \citet{fang2013} and \citet{mypub2015}. Here I just summarize several aspects. The intrinsic color and foreground extinction of a YSO determines its final location in the $JHK_s$ CCD. Figure~\ref{fig:avest}c shows the $H-K_s$ versus $J-H$ CCD of the Class I/II candidates. I split the CCD into three subregions based on the different origins of YSO intrinsic colors. The intrinsic color of a YSO in region 1 is simply assumed to be $[J-H]_0=$\,0.6; the intrinsic color of a YSO in region 2 is obtained by intersecting the reddening vector with the locus of main sequence stars \citep{bb1988}; in region 3, the intrinsic color of a YSO is derived from where the reddening vector and the locus of classical T Tauri stars \citep[CTTS,][]{meyer1997} intersects. The extinction of a YSO \mz{($A_{G,\mathrm{fit}}$)} is calculated by comparing its observed color and intrinsic color with the extinction law suggested by \citet{wang2019}. To estimate the uncertainties of extinctions, I generate a random location in CCD for each Class I/II candidate assuming a normal distribution of its photometric error. Then the extinction value can be obtained based on its location in CCD as mentioned above. I repeat this process ten times for each Class I/II candidate and adopt the mean and standard deviation of ten extinction values as the final value and error of \mz{$A_{G,\mathrm{fit}}$}. Figure~\ref{fig:avest}d shows the de-reddened color $[H-K_s]_0$ versus $[J-H]_0$ CCD for the Class I/II candidates with \mz{$A_{G,\mathrm{fit}}>$}~0 mag.}
    \item[5.]{I combine the remaining sources, including the Class I/II candidates outside the subregions 1,2,3 or without detections in any of $JHK_s$ bands and the Class \mz{III} candidates without \mz{$A_{G}$} or \mz{$A_{G,\mathrm{PNICER}}$}. The extinctions of remaining sources without \mz{$A_{G}$}, \mz{$A_{G,\mathrm{PNICER}}$ or $A_{G,\mathrm{fit}}$} are estimated with the average extinction value of \mz{the} surrounding YSOCs that have extinction measurements in steps 2, 3, and 4. More specifically, the extinctions \mz{($A_{G,\mathrm{avg}}$)} of above remaining sources with distance estimates \mz{($\mathtt{r\_med\_geo}$, see Sect.~\ref{sec:comp-gaiadist})} are obtained by averaging \mz{the neighbours within a radius of 30 pc in 3D space (GLON-GLAT-distance) after excluding outliers with the sigma-clipping technique, and otherwise within a radius of 0.3\degr~in 2D space (GLON-GLAT). Here the values of 30\,pc and 0.3\degr~are the typical radius of the molecular clouds in our Galaxy suggested by \citet{mc2017}.}}
    \item[6.]{I de-redden the SEDs of YSOCs based on their extinction values (\mz{$A_{G,\mathrm{final}}=$~$A_{G}$, $A_{G,\mathrm{PNICER}}$, $A_{G,\mathrm{fit}}$, or $A_{G,\mathrm{avg}}$}) using the extinction law suggested by \citet{wang2019}. Then the de-reddened spectral index, $\alpha_i$ can be obtained, where the subscript $i$ denotes the $i$th iterative loop. Especially, $\alpha_0$ means the observed spectral index. \mz{Then I calculate the difference and corresponding uncertainty of $\alpha_i$ and $A_{G,\mathrm{final},i}$, i.e.,  $\displaystyle \Delta\alpha_i=|\alpha_{i}-\alpha_{i-1}|$, $\displaystyle \sigma(\Delta\alpha_i)=\sqrt{\sigma(\alpha_i)^2+\sigma(\alpha_{i-1})^2}$, $\displaystyle \Delta A_{G,\mathrm{final},i}=|A_{G,\mathrm{final},i}-A_{G,\mathrm{final},i-1}|$, 
    $\sigma(\Delta A_{G,\mathrm{final},i})=\sqrt{\sigma(A_{G,\mathrm{final},i})^2+\sigma(A_{G,\mathrm{final},i-1})^2}$ for each YSOC, where $\alpha_{i}$ and $A_{G,\mathrm{final},i}$ represent the $i$th iterative spetral index and extinction of the YSOC while $\sigma(\alpha_i)$ and $\sigma(A_{G,\mathrm{final},i})$ were their corresponding uncertainties.}}
    \item[7.]{Repeat above steps 2$-$6 until \mz{both $\alpha$ and $A_{G,\mathrm{final}}$ are converged for each YSOC. I define the convergence of $\alpha_i$ and $A_{G,\mathrm{final},i}$ when $\Delta\alpha_i<\sigma(\Delta\alpha_i)$ and $\Delta A_{G,\mathrm{final},i}<\sigma(\Delta A_{G,\mathrm{final},i})$ in $i-2$, $i-1$, and $i$ iterations. }
    Figure~\ref{fig:extloop} shows $\Delta\alpha$ and $\Delta A_{G,\mathrm{final}}$ as a function of iterations \mz{for 26 randomly selected YSOCs. I find that $\alpha$ and $A_{G,\mathrm{final}}$ are converged for $>$90\% YSOCs after 5$-$6 iterations.} 
    Finally I adopt the $7$th iterative spectral indices and extinction values of YSOCs as the de-reddened spectral indices ($\alpha_c$) and extinction estimates.}
\end{itemize}

\begin{figure*}
    \centering
    \includegraphics[width=1.0\linewidth]{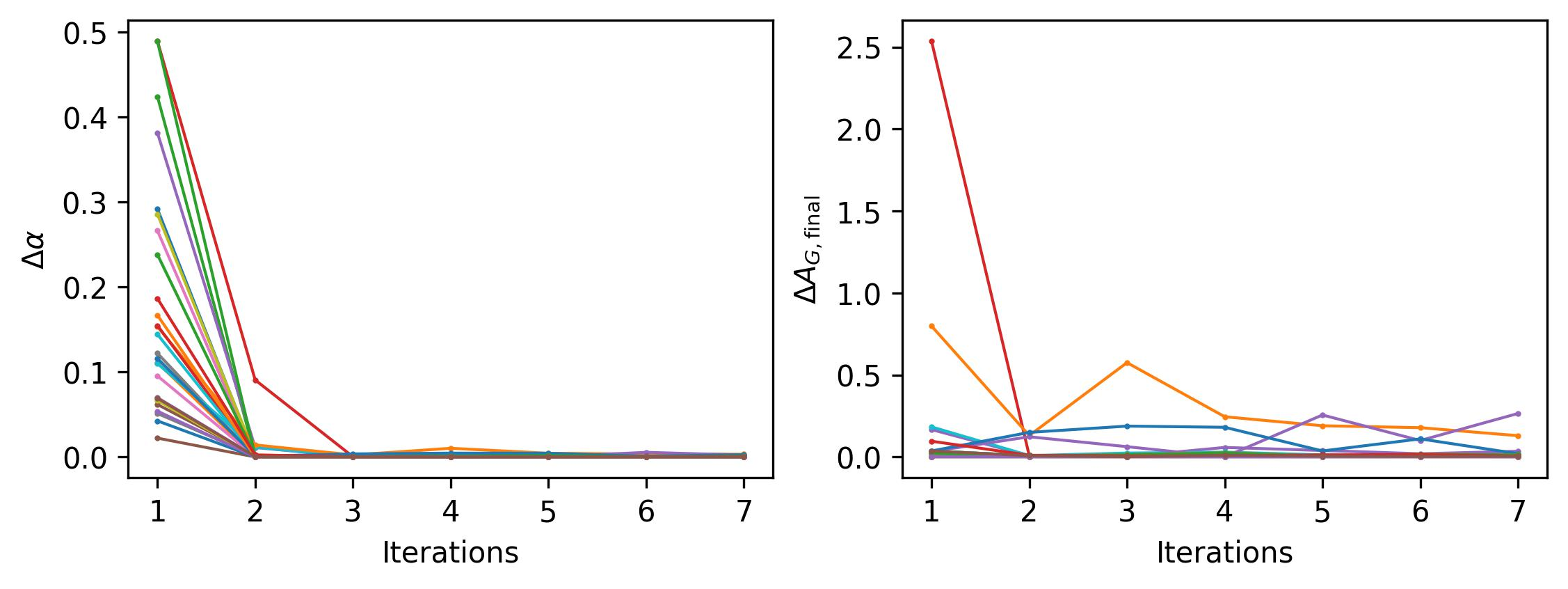}
    \caption{\mz{The $\Delta\alpha$ and $\Delta A_{G,\mathrm{final}}$ of 26 randomly selected YSOCs as a function of iterations.}}
    \label{fig:extloop}
\end{figure*}


\bibliography{ysowise}{}
\bibliographystyle{aasjournal}


\end{CJK*}
\end{document}